\documentclass[prd,aps,showpacs,epsf,floats]{revtex4}%
\usepackage{amssymb}
\usepackage{amsfonts}
\usepackage{amsmath}
\usepackage{graphicx}%
\providecommand{\U}[1]{\protect\rule{.1in}{.1in}}
\begin{document}
\title{\textbf{Geometric Graph-Theoretic Aspects of Quantum Stabilizer Codes}}
\author{\textbf{Carlo Cafaro}}
\affiliation{SUNY Polytechnic Institute, 12203 Albany, New York, USA}

\begin{abstract}
We propose a systematic procedure for the construction of graphs associated
with binary quantum\textbf{ }stabilizer codes. The procedure is characterized
by means of the following three step process. First, the stabilizer code is
realized as a codeword-stabilized (CWS) quantum code. Second, the canonical
form of the CWS\ code is determined and third, the input vertices are attached
to the graphs. In order to verify the effectiveness of the procedure, we
implement the Gottesman stabilizer code characterized by multi-qubit encoding
operators for the resource-efficient error correction of arbitrary
single-qubit errors. Finally, the error-correcting capabilities of the
Gottesman eight-qubit quantum stabilizer code is verified in graph-theoretic
terms as originally advocated by Schlingemann and Werner.

\end{abstract}

\maketitle

\section{Introduction}

We divide the Introduction into two parts. In the first subsection, we discuss
relevant background information. In the second subsection, we present our
motivations together with our main objectives.

\subsection{Background Information}

Classical graphs \cite{die, west, wilson} are known to be intimately related
to quantum error correcting codes (QECCs) \cite{knill97,gotty}. The first
formulation of QECCs constructed from classical graphs and finite dimensional
Abelian groups is attributed to the work of Schlingemann and Werner (SW-work)
\cite{werner}. In Ref. \cite{werner}, the authors pointed out that one problem
with the (non-graphical) schemes of quantum error correction is that the
verification of their error correcting capabilities often requires tedious
computations \cite{cafaro10A,cafaro10B,cafaro11}. They therefore concluded
that it would be desirable to develop new, perhaps more transparent approaches
for constructing error correcting codes, within which more direct geometric
intuition might be exploited. While it was verified in \cite{werner} that all
codes constructed from graphs are stabilizer codes, it nevertheless remains
unclear how typical stabilizer codes may be embedded into the proposed
graphical scheme. For this reason, the full utility of the graphical approach
to quantum coding for stabilizer codes cannot be fully realized unless the
aforementioned embedding issue is resolved. In \cite{dirk}, Schlingemann
(S-work) clarifies this issue by demonstrating that each quantum stabilizer
code (either binary or nonbinary) can be fully described as a graph code and
vice-versa. Almost simultaneously, inspired by the work presented in
\cite{werner}, the equivalence of graphical quantum codes and stabilizer codes
was also uncovered by Grassl et al. in \cite{markus}.

In Refs. \cite{markus,markus1}, the starting point for constructing a graph of
a stabilizer code is the recognition that a stabilizer code corresponds to a
symplectic code viewed as a sub-code of a self-dual code. The construction
begins by selecting a generator matrix $G_{\text{old}}^{\prime}\overset
{\text{def}}{=}\left(  X^{\prime}|Z^{\prime}\right)  $ for the self-dual code.
Then, with a clever joint action on $G_{\text{old}}^{\prime}$ by the symmetric
group (permutation of the columns of the matrix), the symplectic group (local
action on sub-matrices of the generator matrix), and by additional column
operations, $G_{\text{old}}^{\prime}$ can be recast as $G_{\text{new}}%
^{\prime}\overset{\text{def}}{=}\left(  I|C\right)  $ with $C$ being a
symmetric matrix with all entries on the diagonal of $C$ equal to zero. The
code generated by $G_{\text{new}}^{\prime}$ is equivalent to the self-dual
code. Finally, a graphical QECC which is equivalent to the stabilizer code
corresponding to the symplectic code is specified by the adjacency matrix
$\Gamma\overset{\text{def}}{=}\left[  O\text{, }B\text{; }B^{t}\text{,
}C\right]  $ with $B^{t}$ defining the symplectic code as a sub-code of the
self-dual code. Despite its undeniable generality, the scheme presented by
Grassl et \textit{al}. does not address in an explicit, practical manner how
to construct the self-dual code in terms of its generator matrix. Furthermore,
the scheme does not explain in a specific manner how to transition from
$G_{\text{old}}^{\prime}$ to $G_{\text{new}}^{\prime}$ via the wreath product
of the symplectic group $\mathrm{Sp}_{2}\left(  p\right)  $ and the symmetric
group $\mathrm{S}_{n}$ (that is, the group of isometries of the symplectic
space $\mathbf{F}_{p^{2}}^{n}$ that additionally preserve the Hamming weight).

A crucial advancement for the description and understanding of the connections
among properties of graphs and stabilizer codes was achieved due to the
introduction of graph states (and cluster states \cite{hans}) in the graphical
construction of QECCs as presented by Hein et \textit{al}. in \cite{hein}. In
this work the correspondence between graph states and graphs was shown and
special attention was devoted to understanding how the entanglement in a graph
state is related to the topology of its underlying graph. In \cite{hein}, it
was also emphasized that codewords of various QECCs can be regarded as special
instances of graph states. Furthermore, criteria for the equivalence of graph
states under local unitary transformations expressed entirely on the level of
the underlying graphs were presented. Similar findings were deduced by Van den
Nest et \textit{al}. in \cite{bart} (VdN-work) where a constructive scheme
that establishes how each stabilizer state is equivalent to a graph state
under the action of local Clifford operations was presented. In particular, an
algorithmic procedure for transforming any binary quantum stabilizer code into
a graph code was described. In this manner, the primary finding of
Schlingemann in \cite{dirk} is reproduced in \cite{bart} for the special case
of binary quantum states. Most significantly, an algorithmic procedure for
transforming any binary quantum stabilizer code into a graph code appears in
\cite{bart}. To the best of our knowledge, the results provided by
Schlingemann in \cite{dirk} and Van den Nest et al. in \cite{bart}
have\textbf{ }yet to be fully and jointly exploited to establish a systematic
procedure for generating graphs associated with arbitrary binary stabilizer
codes with emphasis on verification of their error-correcting capabilities.
Indeed, this last point represents one of the original motivations for
introducing graphs into quantum error correction (QEC) \cite{werner}.

The codeword-stabilized (CWS) quantum code formalism represents a unifying
scheme for the construction of either additive or nonadditive QECCs, for cases
involving both binary \cite{cross} (CWS-work) and nonbinary states
\cite{chen}. Furthermore, every CWS code in its canonical form can be
completely characterized by means of a graph and a classical code. In
particular, any CWS code is locally Clifford equivalent to a CWS code with a
graph state stabilizer and word operators consisting only of $Z$s
\cite{cross}. Since stabilizer codes, graph codes and graph states can all be
recast within the CWS formalism, it should prove instructive to investigate
the graphical depiction of stabilizer codes as originally envisioned by
Schlingemann and Werner within the context of the generalized framework
wherein stabilizer codes are realized as CWS\ codes. Proceeding along this
path of inquiry, it will become evident that graph states in QEC as presented
in \cite{hein} emerges naturally. What is more, the algorithmic procedure for
transforming any binary quantum stabilizer code into a graph code as depicted
in \cite{bart} can be employed and jointly leveraged with the results obtained
in \cite{dirk}, where the notions of coincidence and adjacency matrices of
classical graphs in QEC are introduced. For the sake of completeness, we
emphasize that the CWS formalism has already been used in the literature for
the graphical construction of binary \cite{yu1} as well as nonbinary
\cite{yu2} (both additive/stabilizer and nonadditive) QECCs. In \cite{yu1} for
instance, by regarding stabilizer codes as CWS codes and utilizing a graphical
approach to quantum coding, a classification of all extremal stabilizer codes
up to eight qubits along with the formulation of the optimal $\left(  \left(
10\text{, }24\text{, }3\right)  \right)  $ code together with a family of
$1$-error detecting nonadditive codes exhibiting the highest encoding rate, so
far obtained, was presented. With the foresight of imagination, a graphical
quantum computation based directly on graphical objects is also envisioned in
\cite{yu1}. Indeed, this vision recently moved a step closer to reality in the
work of Beigi et \textit{al}. \cite{beigi}. In this article, a systematic
procedure for generating binary and non-binary concatenated quantum codes
based on graph concatenation (essentially) within the CWS framework was
developed. Graphs corresponding to both the inner and the outer codes are
concatenated by means of a simple graph operation, namely the generalized
local complementation operation \cite{beigi}. Despite their findings, the
authors of \cite{beigi} correctly point out that the elusive role played by
graphs in QEC is still poorly understood. Indeed, in neither \cite{yu1} nor
\cite{beigi} are the authors concerned with the joint exploitation of the
results provided by Van den Nest et \textit{al.} in \cite{bart} or
Schlingemann in \cite{dirk} to provide a systematic procedure for constructing
graphs associated with arbitrary binary stabilizer codes with emphasis on the
verification of their error correcting capabilities. By contrast, we aim in
the present work to investigate such topics and further, hope to enhance our
understanding of the role played by classical graphs in quantum coding.

\subsection{Motivations and Objectives}

In this article, we propose a systematic scheme for the construction of graphs
exhibiting both input and output vertices associated with arbitrary binary
stabilizer codes. Our scheme is implemented via the following three step
process: first, the stabilizer code is realized as a CWS quantum code; second,
the canonical form of the CWS\ code is determined; third, the input vertices
are attached to those graphs initially containing only output vertices. In
order to verify the effectiveness of our scheme, we present the graphical
construction of the Gottesman $\left[  \left[  8\text{, }3\text{, }3\right]
\right]  $ stabilizer code characterized by multi-qubit encoding operators
\cite{danielpra}. In particular, the error-correcting capabilities of the
$\left[  \left[  8\text{, }3\text{, }3\right]  \right]  $ eight-qubit quantum
stabilizer code is verified in graph-theoretic terms as originally advocated
by Schlingemann and Werner. Finally, possible generalizations of our scheme
for the graphical construction of both stabilizer and nonadditive nonbinary
quantum codes is briefly addressed.

Our proposed scheme differs from the\textbf{ }Grassl et \emph{al}. scheme
\cite{markus,markus1} in several respects.

\begin{itemize}
\item \emph{Different motivations}. The two schemes were proposed to address
different sets of motivations. The 2002 Grassl et \textit{al}. work was
motivated by the desire to provide an answer to the question raised by
Schlingemann and Werner in 2001 on whether or not every stabilizer code was
equivalent to a graphical quantum error correction code. By contrast, our
investigation is partially motivated by the scientific curiosity to re-examine
the graphical quantum error correction conditions proposed by Schlingemann and
Werner in 2001 in the presence of the established connection between graph
codes and stabilizer codes by Grassl et \textit{al}. and, independently, by
Schlingemann in 2002.

\item \emph{Different range of applicability}. The two schemes have different
fields of relevance. Our scheme is less general and abstract. In particular,
unlike the Grassl et \textit{al}. scheme, our method is limited to two-level
quantum systems (qubits) and binary stabilizer codes. Furthermore, we do not
consider in a quantitative manner $d$-level quantum systems (qudits) and
nonbinary stabilizer codes. Despite its lack of generality, our scheme does
apply in a very practical manner to any binary stabilizer code. Moreover, it
does provide a simple, explicit method for analytically constructing a
graphical depiction of a stabilizer code with graph-theoretic verification of
its error correction capabilities. To the best of our knowledge, a scheme with
these particular features is absent in the literature.

\item \emph{Different techniques}. The two schemes utilize different
techniques. In the Grassl et \textit{al}. scheme, it is explained that the
construction of a graphical representation of a stabilizer code can be
achieved in three steps: first, the scheme exploits the fact that any
stabilizer code that is specified via a code over $\mathbf{F}_{p^{m}}$ can be
recast as a stabilizer code over the prime field $\mathbf{F}_{p}$ (that is,
symplectic codes over $\mathbf{F}_{p^{2}}$ \cite{knill01}); second, the
stabilizer corresponding to a graphical code is computed by means of finite
symplectic geometry methods \cite{rains99}; finally, after suitable matrix
computations, the construction of a graphical representation of a stabilizer
code follows. In particular, the technique used in their work establishes a
connection between quadratic forms and quantum codes. Our scheme, as
previously mentioned, is obtained by combining, in a suitably chosen logical
manner, previously known results: i) the 2009 CWS formalism by Cross et
\textit{al}. \cite{cross}; ii) the 2004 equivalence of graph states under
local unitary Clifford transformations by Van den Nest et \textit{al}.
\cite{bart}; iii) the 2002 connection between graphs and stabilizer codes
\cite{dirk}. These techniques employed in our scheme were clearly not
available to Grassl et \textit{al}. in 2002. While quadratic forms and their
connection to graphical quantum codes play a key, explicit role in the Grassl
et \textit{al}. scheme, they do not in our approach. The role played by
quadratic forms is covered by graph states in our scheme. In addition, the
role played by transformations in symmetric and symplectic groups is replaced
in our work by local unitary and local Clifford transformations. In addition,
while Grassl et \textit{al}. work was primarily conceptual, our work is more
applied in nature. Despite our practical original motivation however, in the
process of implementing our scheme we establish intriguing connections among
the CWS formalism, the algorithmic procedure to transform a stabilizer state
into a graph state, and the work on the link between graph and stabilizer codes.
\end{itemize}

For completeness, we remark that this paper is a rewritten shorter version of
the unpublished work in Ref. \cite{graph14}. In this shorter version, we limit
our discussion to stabilizer binary codes, we chose to discuss in detail a
single application of our proposed scheme and, above all, we clarify our
objectives by underlying the differences between our scheme and existing
previous works on graphical constructions of quantum stabilizer codes. In this
way, we hope to achieve a higher degree of clarity and simplicity so that our
work can be fully appreciated by the interested readers.

The layout of this article is as follows. In Section II, the CWS quantum code
formalism is explained in brief. In Section III, we re-examine several basic
elements of the Schlingemann-Werner work (SW-work, \cite{werner}), the
Schlingemann work (S-work, \cite{dirk}) and, finally, the Van den Nest et al.
work (VdN-work, \cite{bart}). We focus on the aspects of these works that are
relevant to the development of our scheme. In Section IV, we formally describe
our scheme and, as an illustrative prominent and important example, apply it
to the graphical construction of the Gottesman $\left[  \left[  8\text{,
}3\text{, }3\right]  \right]  $ quantum stabilizer code
\cite{danielpra,cafaro14a,cafaro14b}. Concluding remarks appear in Section V.
Finally, we present some preliminary technical material on graphs, graph
states, graph codes, local Clifford transformations on graph states as well as
local complementations on graphs in Appendix A.

\section{The Codeword-Stabilized work}

As mentioned in the Introduction, we propose an alternative, systematic
approach to find a graphical depiction of binary stabilizer quantum codes. Our
scheme provides a methodical and explicit step-by-step analytic construction
of graphical depictions of stabilizer codes.\textbf{ }Our method, restricted
to binary stabilizer codes, can be regarded as an alternative to the 2002
Grassl et \textit{al}. scheme. In order to facilitate an ease of reading, we
refer to preliminary aspects of graphs, graph states, graph codes, local
Clifford transformations on graph states as well as local complementations on
graphs in Appendix A. In what follows, we begin by focusing on the CWS-work.

The category of CWS codes includes the set of all stabilizer codes as well as
several nonadditive codes. For the sake of completeness, we point out the
existence of quantum codes that cannot be recast within the CWS framework as
mentioned in \cite{cross} and illustrated in \cite{ruskai}. CWS codes cast in
standard form can be specified in terms of a graph $G$ and a (nonadditive, in
general) classical binary code $\mathcal{C}_{^{\text{classical}}}$. The $n$
vertices of the graph $G$ correspond to the $n$ qubits of the code, while the
adjacency matrix of graph\textbf{ }$G$ is denoted $\Gamma$. Given the graph
state $\left\vert G\right\rangle $ and the binary code $\mathcal{C}%
_{^{\text{classical}}}$, a unique base state $\left\vert S\right\rangle $
along with a set of word operators $\left\{  w_{l}\right\}  $ can be
specified. The base state $\left\vert S\right\rangle $ is a single stabilizer
state that is stabilized by the word stabilizer $\mathcal{S}_{\text{CWS}}$,
with the latter being an element of a maximal Abelian subgroup of the Pauli
group $\mathcal{P}_{\mathcal{H}_{2}^{n}}$.

Let us denote by\textbf{ }$\left(  \left(  n\text{, }K\text{, }d\right)
\right)  $ a quantum code of $n$ qubits that serves to encode $K$ dimensions
with distance $d$. Following \cite{cross}, it can be shown that an $\left(
\left(  n\text{, }K\text{, }d\right)  \right)  $ codeword stabilized code with
word operators $\mathcal{W}\overset{\text{def}}{=}\left\{  w_{l}\right\}  $
where $l\in\left\{  1\text{,..., }K\right\}  $ and codeword stabilizer
$\mathcal{S}_{\text{CWS}}$ is locally Clifford equivalent to a\textbf{
}codeword stabilized code with word operators $\mathcal{W}^{\prime}$,%
\begin{equation}
\mathcal{W}^{\prime}\overset{\text{def}}{=}\left\{  w_{l}^{\prime
}=Z^{\mathbf{c}_{l}}\right\}  \text{,} \label{can1}%
\end{equation}
and codeword stabilizer $\mathcal{S}_{\text{CWS}}^{\prime}$,%
\begin{equation}
\mathcal{S}_{\text{CWS}}^{\prime}\overset{\text{def}}{=}\left\langle
S_{l}^{\prime}\right\rangle =\left\langle X_{l}Z^{\mathbf{r}_{l}}\right\rangle
\text{,} \label{can2}%
\end{equation}
where the quantities $\mathbf{c}_{l}$s are codewords defining the classical
binary code $\mathcal{C}_{\text{classical}}$ and $\mathbf{r}_{l}$ is the $l$th
row vector of the adjacency matrix $\Gamma$ of the graph $G$. Observe that
$I^{i}$, $X^{i}$, $Y^{i}$, $Z^{i}$ (or, equivalently, $X^{i}=\sigma_{x}^{i}$,
$Y^{i}=\sigma_{y}^{i}$, $Z^{i}=\sigma_{z}^{i}$) denote the identity matrix and
the three Pauli operators acting on the qubit $i\in V$ with $V$ denoting the
set of vertices of the graph. For clarity, we point out that the quantity
$Z^{\mathbf{v}}$ appearing\textbf{ }in Eq. (\ref{can2}) is the notational
shorthand for%
\begin{equation}
Z^{\mathbf{v}}\overset{\text{def}}{=}Z^{v_{1}}\otimes\text{...}\otimes
Z^{v_{n}}\text{,}%
\end{equation}
where the symbol \textquotedblleft$\otimes$\textquotedblright\ denotes the
tensor product (which, for notational simplicity, may be omitted in the rest
of the manuscript) and $\mathbf{v}\overset{\text{def}}{=}\left(
v_{1}\text{,..., }v_{n}\right)  \in\mathbf{F}_{2}^{n}$ is a binary
$n$-dimensional vector with $\mathbf{F}_{2}\overset{\text{def}}{=}\left\{
0\text{, }1\right\}  $. It is therefore evident that\textbf{ }any CWS code is
locally Clifford equivalent to a CWS code with a graph-state stabilizer and
word operators consisting only of $Z$s. Furthermore, the word operators can
always be chosen so as to include the identity. It can be said that Eqs.
(\ref{can1})\ and (\ref{can2}) effectively characterize the so-called standard
form of a CWS quantum code. For a CWS code expressed in standard form, the
base state $\left\vert S\right\rangle $ is a graph state. Furthermore, the
code-space of a CWS\ code is spanned by a set of basis vectors resulting from
application of the word operators $w_{l}$ on the base state $\left\vert
S\right\rangle $, yielding%
\begin{equation}
\mathcal{C}_{\text{CWS}}\overset{\text{def}}{=}\text{\textrm{Span}}\left\{
\left\vert w_{l}\right\rangle \right\}  \text{ with, }\left\vert
w_{l}\right\rangle \overset{\text{def}}{=}w_{l}\left\vert S\right\rangle
\text{.}%
\end{equation}
Hence, the dimension of the code-space is equal to the number of word
operators. These operators are Pauli operators belonging to $\mathcal{P}%
_{\mathcal{H}_{2}^{n}}$ that anti-commute with one or more of the stabilizer
generators of the base state. Hence, word operators act to map the base state
onto an orthogonal state. The only exception is, in general, the inclusion of
the identity operator within the set of word operators. This fact implies that
the base state is also a codeword of the quantum code. Indeed, these base
states are simultaneously eigenstates of the stabilizer generators with the
exception that some of the eigenvalues differ from $+1$. Additionally, it can
be verified that a single qubit Pauli error $X$, $Z$ or $ZX$ acting on a
codeword $\omega\left\vert S\right\rangle $ of a CWS code in standard form is
equivalent (up to a sign) to another multi-qubit error consisting of $Z$s with
$\omega$ denoting a word operator. Since all errors reduce to $Z$s, the
original quantum error model can be transformed into a classical error model
characterized, in general, by multi-qubit errors. The map $\mathcal{C}%
l_{\mathcal{S}_{\text{CWS}}}$ that defines this transformation reads,%
\begin{equation}
\mathcal{C}l_{\mathcal{S}_{\text{CWS}}}:\mathcal{E}\ni E\overset{\text{def}%
}{=}\pm Z^{\mathbf{v}}X^{\mathbf{u}}\mapsto\mathcal{C}l_{\mathcal{S}%
_{\text{CWS}}}\left(  \pm Z^{\mathbf{v}}X^{\mathbf{u}}\right)  \overset
{\text{def}}{=}\mathbf{v\oplus}%
%TCIMACRO{\dbigoplus \limits_{l=1}^{n}}%
%BeginExpansion
{\displaystyle\bigoplus\limits_{l=1}^{n}}
%EndExpansion
u_{l}\mathbf{r}_{l}\in\left\{  0\text{, }1\right\}  ^{n}\text{,}%
\end{equation}
where $\mathcal{E}$ denotes the set of Pauli errors $E$, $\mathbf{r}_{l}$ is
the $l^{\text{th}}$\textit{\ }row of the adjacency matrix $\Gamma$ for the
graph $G$ and $u_{l}$ is the $l$\textit{th }bit of the vector $\mathbf{u}$. At
this juncture, we point out that it was demonstrated in \cite{cross} that any
stabilizer code is a CWS code. Specifically, a quantum stabilizer code
$\left[  \left[  n,k,d\right]  \right]  $ (where the parameters $n$, $k$, $d$
denote the length, dimension and distance of the quantum code, respectively)
with stabilizer $\mathcal{S}\overset{\text{def}}{=}\left\langle S_{1}%
\text{,..., }S_{n-k}\right\rangle $ where $S_{j}$ with $j\in\left\{
1\text{,..., }n-k\right\}  $ denote the stabilizer generators $\bar{X}_{1}%
$,..., $\bar{X}_{k}$ and logical operations $\bar{Z}_{1}$,..., $\bar{Z}_{k}$
is equivalent to a CWS code defined by,%
\begin{equation}
\mathcal{S}_{\text{CWS}}\overset{\text{def}}{=}\left\langle S_{1}\text{, ...,
}S_{n-k}\text{, }\bar{Z}_{1}\text{,..., }\bar{Z}_{k}\right\rangle \text{,}%
\end{equation}
and word operators $\omega_{\mathbf{v}}$,%
\begin{equation}
\omega_{\mathbf{v}}\overset{\text{def}}{=}\bar{X}_{1}^{\left(  \mathbf{v}%
\right)  _{1}}\otimes\text{...}\otimes\bar{X}_{k}^{\left(  \mathbf{v}\right)
_{k}}\text{.}%
\end{equation}
Note that the vector $\mathbf{v}$ denotes a $k$-bit string and $\left(
\mathbf{v}\right)  _{l}\overset{\text{def}}{=}v_{l}$ with $l\in\left\{
1\text{,..., }k\right\}  $ is the $l^{\text{th}}$ bit of the vector
$\mathbf{v}$. For further details on binary CWS\ quantum codes, we direct the
reader to \cite{cross}. Finally, for a recent investigation on the symmetries
of CWS codes, we refer to \cite{tqc}.

\section{From graphs to stabilizer codes and vice-versa}

In this Section, we turn our attention to a re-examination of some basic
elements of the Schlingemann-Werner work (SW-work, \cite{werner}), the
Schlingemann work (S-work, \cite{dirk}) and, finally, the Van den Nest et
\textit{al}. work (VdN-work, \cite{bart}). We restrict our focus to the
aspects of these works that are especially relevant for the implementation of
our proposed scheme.

\subsection{The Schlingemann-Werner work}

The basic graphical formulation of quantum codes within the SW-work
\cite{werner} can be described as follows. Quantum codes can be completely
defined in terms of a uni-directed graph $G\left(  V\text{, }E\right)  $ which
is\textbf{ }characterized by a set $V$ of $n$ vertices and a set of edges $E$,
specified by a coincidence matrix $\Xi$, exhibiting input and output vertices
as well as a finite Abelian group structure\textbf{ }$\mathcal{G}$ with
non-degenerate, symmetric bi-character $\chi$. We remark that there are
various types of matrices that may be used to specify a given graph such as
incidence and adjacency matrices \cite{die}, for instance. The coincidence
matrix introduced in \cite{werner} is essentially the adjacency matrix of a
graph with both input and output vertices. The adjacency matrix\textbf{
}should not be confused with the so-called incidence matrix of a graph. The
sets of input and output vertices are denoted by $X$ and $Y$, respectively.
Let $\mathcal{G}$ represent any finite, additive Abelian group of cardinality
$\left\vert \mathcal{G}\right\vert =n$, with the addition operation denoted by
"$+$" and null element $0$. A non-degenerate symmetric bi-character is a map,
\begin{equation}
\chi:\mathcal{G}\times\mathcal{G}\ni\left(  g\text{, }h\right)  \mapsto
\chi\left(  g\text{, }h\right)  =\left\langle g\text{, }h\right\rangle \in%
%TCIMACRO{\U{2102}}%
%BeginExpansion
\mathbb{C}%
%EndExpansion
\end{equation}
satisfying the following properties: (i) $\left\langle g\text{, }%
h\right\rangle =\left\langle h\text{, }g\right\rangle $, $\forall g$,
$h\in\mathcal{G}$; (ii) $\left\langle g\text{, }h_{1}+h_{2}\right\rangle
=\left\langle g\text{, }h_{1}\right\rangle \left\langle g\text{, }%
h_{2}\right\rangle $, $\forall g$, $h_{1}$, $h_{2}\in\mathcal{G}$; (iii)
$\left\langle g\text{, }h\right\rangle =1$ $\forall h\in\mathcal{G}%
\Leftrightarrow g=0$. If $\mathcal{G}=%
%TCIMACRO{\U{2124} }%
%BeginExpansion
\mathbb{Z}
%EndExpansion
_{n}\overset{\text{def}}{=}\left\{  0\text{,..., }n-1\right\}  $ (the cyclic
group of order $n$) with addition modulo $n$ specifies a group operation, then
the bi-character map\textbf{ }$\chi$\textbf{ }may be chosen as%
\begin{equation}
\chi\left(  g\text{, }h\right)  =\left\langle g\text{, }h\right\rangle
\overset{\text{def}}{=}e^{i\frac{2\pi}{n}gh}\text{, }%
\end{equation}
with $g$, $h\in%
%TCIMACRO{\U{2124} }%
%BeginExpansion
\mathbb{Z}
%EndExpansion
_{n}$. The encoding operator $\mathbf{v}_{G}$ of an error correcting code is
an isometry (i.e. a bijective map between two metric spaces that preserve
distances),%
\begin{equation}
\mathbf{v}_{G}:L^{2}\left(  \mathcal{G}^{X}\right)  \rightarrow L^{2}\left(
\mathcal{G}^{Y}\right)  \text{,}%
\end{equation}
where $L^{2}\left(  \mathcal{G}^{X}\right)  $ is the $\left\vert X\right\vert
$-fold tensor product $\mathcal{H}^{\otimes X}$ with $\mathcal{H}=L^{2}\left(
\mathcal{G}\right)  $. The Hilbert space $\mathcal{H}$ is the space of
integrable functions over $\mathcal{G}$ with $\mathcal{G}\overset{\text{def}%
}{=}%
%TCIMACRO{\U{2124} }%
%BeginExpansion
\mathbb{Z}
%EndExpansion
_{2}$ in the qubit case. Similarly, $L^{2}\left(  \mathcal{G}^{Y}\right)  $ is
the $\left\vert Y\right\vert $-fold tensor product $\mathcal{H}^{\otimes Y}$.
The Hilbert space $L^{2}\left(  \mathcal{G}\right)  $ is defined as,%
\begin{equation}
L^{2}\left(  \mathcal{G}\right)  \overset{\text{def}}{=}\left\{
\psi\left\vert \psi:\mathcal{G}\rightarrow%
%TCIMACRO{\U{2102} }%
%BeginExpansion
\mathbb{C}
%EndExpansion
\right.  \right\}  \text{,}%
\end{equation}
with the scalar product between two elements $\psi_{1}$ and $\psi_{2}$ in
$L^{2}\left(  \mathcal{G}\right)  $ being defined by,%
\begin{equation}
\left\langle \psi_{1}\text{, }\psi_{2}\right\rangle \overset{\text{def}}%
{=}\frac{1}{\left\vert \mathcal{G}\right\vert }\sum_{g}\bar{\psi}_{1}\left(
g\right)  \psi_{2}\left(  g\right)  \text{.}%
\end{equation}
The action of $\mathbf{v}_{G}$ on $L^{2}\left(  \mathcal{G}^{X}\right)  $ is
prescribed by \cite{werner},%
\begin{equation}
\left(  \mathbf{v}_{G}\psi\right)  \left(  g^{Y}\right)  \overset{\text{def}%
}{=}\int dg^{X}\mathbf{v}_{G}\left[  g^{X\cup Y}\right]  \psi\left(
g^{X}\right)  \text{,} \label{g12}%
\end{equation}
where $\mathbf{v}_{_{G}}\left[  g^{X\cup Y}\right]  $ represents the integral
kernel of isometry $\mathbf{v}_{_{G}}$ and is given by \cite{werner},%
\begin{align}
\mathbf{v}_{G}\left[  g^{X\cup Y}\right]   &  =\left\vert \mathcal{G}%
\right\vert ^{\frac{\left\vert X\right\vert }{2}}\prod\limits_{\left\{
z\text{, }z^{\prime}\right\}  }\chi\left(  g_{z}\text{, }g_{z^{\prime}%
}\right)  ^{\Xi\left(  z\text{, }z^{\prime}\right)  }=\left\vert
\mathcal{G}\right\vert ^{\frac{\left\vert X\right\vert }{2}}\prod
\limits_{\left\{  z\text{, }z^{\prime}\right\}  }\left[  \exp\left(
\frac{2\pi i}{p}g_{z}g_{z^{\prime}}\right)  \right]  ^{\Xi\left(  z\text{,
}z^{\prime}\right)  }\nonumber\\
& \nonumber\\
&  =\left\vert \mathcal{G}\right\vert ^{\frac{\left\vert X\right\vert }{2}%
}\prod\limits_{\left\{  z\text{, }z^{\prime}\right\}  }\left[  \exp\left(
\frac{2\pi i}{p}g_{z}\Xi\left(  z\text{, }z^{\prime}\right)  g_{z^{\prime}%
}\right)  \right]  =\left\vert \mathcal{G}\right\vert ^{\frac{\left\vert
X\right\vert }{2}}\exp\left(  \frac{\pi i}{p}g^{X\cup Y}\cdot\Xi\cdot g^{X\cup
Y}\right)  \text{.} \label{g11}%
\end{align}
The product in Eq. (\ref{g11}) must be taken over each two-element subsets
$\left\{  z\text{, }z^{\prime}\right\}  $ in $X\cup Y$. By\textbf{
}substituting Eq. (\ref{g11}) into Eq. (\ref{g12}), the action of
$\mathbf{v}_{G}$ upon $L^{2}\left(  \mathcal{G}^{X}\right)  $ takes the form,%
\begin{equation}
\left(  \mathbf{v}_{G}\psi\right)  \left(  g^{Y}\right)  =\int dg^{X}%
\left\vert \mathcal{G}\right\vert ^{\frac{\left\vert X\right\vert }{2}}%
\exp\left(  \frac{\pi i}{p}g^{X\cup Y}\cdot\Xi\cdot g^{X\cup Y}\right)
\psi\left(  g^{X}\right)  \text{.} \label{g13}%
\end{equation}
Recall that the sequential steps of a QEC cycle can be described as follows,%
\begin{equation}
\rho\overset{\text{coding}}{\longrightarrow}\mathbf{v}\rho\mathbf{v}^{\ast
}\equiv\rho^{\prime}\text{, }\rho^{\prime}\overset{\text{noise}}%
{\longrightarrow}\mathbf{T}\left(  \rho^{\prime}\right)  =\sum\limits_{\alpha
}F_{\alpha}\rho^{\prime}F_{\alpha}^{\ast}\equiv\rho^{\prime\prime}\text{,
}\rho^{\prime\prime}\overset{\text{recovery}}{\longrightarrow}\mathbf{R}%
\left(  \rho^{\prime\prime}\right)  =\rho\text{,}%
\end{equation}
that is,%
\begin{equation}
\mathbf{R}\left(  \mathbf{T}\left(  \mathbf{v}\rho\mathbf{v}^{\ast}\right)
\right)  =\rho\text{,} \label{HC}%
\end{equation}
where the symbol \textquotedblleft$\ast$\textquotedblright\ in Eq. (\ref{HC})
denotes Hermitian conjugation. Furthermore, the traditional Knill-Laflamme
error-correction conditions read,%
\begin{equation}
\left\langle \mathbf{v}\psi_{1}\text{, }F_{\alpha}^{\ast}F_{\beta}%
\mathbf{v}\psi_{2}\right\rangle =\omega\left(  F_{\alpha}^{\ast}F_{\beta
}\right)  \left\langle \psi_{1}\text{, }\psi_{2}\right\rangle \text{,}
\label{kl}%
\end{equation}
where the multiplicative factor $\omega\left(  F_{\alpha}^{\ast}F_{\beta
}\right)  $ is independent of the states $\psi_{1}$ and $\psi_{2}$. The
graphical analogue of Eq. (\ref{kl}) is given by,%
\begin{equation}
\left\langle \mathbf{v}\psi_{1}\text{, }F\mathbf{v}\psi_{2}\right\rangle
=\omega\left(  F\right)  \left\langle \psi_{1}\text{, }\psi_{2}\right\rangle
\text{,} \label{g14}%
\end{equation}
for all operators $F$ in $\mathcal{U}(E)$\textbf{\ }which coincides
with\textbf{ }the set of operators in $L^{2}(\mathcal{G}^{Y})$ that are
localized in $E\subset Y$. Hence, operators in $\mathcal{U}(E)$ are given by
the tensor product of an arbitrary operator on $\mathcal{H}^{\otimes E}$ with
the identity on $\mathcal{H}^{\otimes Y\backslash E}$. Note that $\left\{
F_{\alpha}\right\}  $ in Eq. (\ref{kl}) are error operators that belong to a
linear subspace of operators acting on the output Hilbert space. The
quantity\textbf{ }$F$\textbf{ }in Eq. (\ref{g14}), instead, is an element of
$\mathcal{U}(E)$\textbf{. }Following Ref. \cite{werner}, the transition from
Eq. (\ref{kl}) to Eq. (\ref{g14}) occurs once one realizes that the error
operators\textbf{ }$F_{\alpha}^{\ast}F_{\beta}$ in\ Eq. (\ref{kl}) can be
localized on arbitrary sets of $2e$\textbf{ }elements. Thus, any operator
which exhibits this localization property can be recast as a linear
combination of such errors\textbf{ }$F_{\alpha}^{\ast}F_{\beta}$\textbf{.}

A graph code corrects $e$ errors provided it detects all error configurations
$E\subset Y$ with $\left\vert E\right\vert \leq2e$. Given such a graphical
construction of the encoding operator $\mathbf{v}_{G}$ as appears\textbf{ }in
Eq. (\ref{g13}), together with the graphical quantum error-correction
conditions in Eq. (\ref{g14}), the primary finding of Schlingemann and Werner
can be restated as follows: given a finite Abelian group $\mathcal{G}$ and a
weighted graph $G$, an error configuration $E\subset Y$ is detected by the
quantum code $\mathbf{v}_{G}$ if and only if given%
\begin{equation}
d^{X}=0\text{ and }\Xi_{E}^{X}d^{E}=0\text{,} \label{wc1}%
\end{equation}
then,%
\begin{equation}
\Xi_{X\cup E}^{I}d^{X\cup E}=0\Rightarrow d^{X\cup E}=0\text{,} \label{wc2}%
\end{equation}
with $I=Y\backslash E$. In general, the condition\textbf{\ }$\Xi_{B}^{A}%
d^{B}=0$\ constitute a set of equations, one for each integration
vertex\textbf{\ }$a\in A$\textbf{: }for each vertex\textbf{\ }$a\in
A$\textbf{, }we must sum $d_{b}$\textbf{\ }for all vertices\textbf{\ }$b\in
B$\textbf{\ }connected to\textbf{\ }$a$\textbf{, }and equate it to
zero\textbf{. }Furthermore, we emphasize that the fact $v_{G}$\ is
an\textbf{\ }isometry is equivalent to the detection of zero errors. As
pointed out by Werner and Schlingemann in Ref. \cite{werner}, this can be can
be explained by setting $F=\mathbf{1}$\textbf{ }in Eq. (\ref{g14})\textbf{.
}In graph-theoretic terms, the detection of zero errors requires $\Xi_{X}%
^{Y}d^{X}=0$,\textbf{\ }implying that\textbf{\ }$d^{X}=0$. We assume that Eq.
(\ref{wc2}) with the additional constraints in Eq. (\ref{wc1}) constitute the
weak version (necessary and sufficient conditions) of the graph-theoretic
error detection conditions. It is worth noting that sufficient (but not
necessary) graph-theoretic error detection conditions can be introduced as
well. In particular, an error configuration $E$ is detectable by a quantum
code if,%
\begin{equation}
\Xi_{X\cup E}^{I}d^{X\cup E}=0\Rightarrow d^{X\cup E}=0\text{.} \label{sc}%
\end{equation}
We refer to the conditions in Eq. (\ref{sc}) without any additional
graph-theoretic constraints like those provided in Eq. (\ref{wc1}) as the
strong version (sufficient conditions) of the graph-theoretic error detection
conditions. For completeness, we emphasize that the non-graphical sufficient
but not necessary (that is, strong) error correction conditions are specified
by the relations%
\begin{equation}
\left\langle \mathbf{v}\psi_{1}\text{, }F_{\alpha}^{\ast}F_{\beta}%
\mathbf{v}\psi_{2}\right\rangle =\delta_{\alpha\beta}\left\langle \psi
_{1}\text{, }\psi_{2}\right\rangle \text{,}%
\end{equation}
while the necessary and sufficient (that is, weak) Knill-Laflamme error
correction conditions are in\ Eq. (\ref{kl}). Furthermore, we recall that in
graph-theoretic terms, a code corrects $e$ errors if and only if it detects
all error configurations $E\subset Y$ with $\left\vert E\right\vert \leq2e$.
Therefore, the ability to correct any one-qubit error demands the
detectability of any two-qubit error. In this sense, we may interchangeably
talk about either error correction or error detection conditions in our work.

\subsection{The Schlingemann-work}

Schlingemann was able to demonstrate that stabilizer codes, either binary or
nonbinary, are equivalent to graph codes and vice-versa. In the context
of\textbf{ }our proposed scheme, the main finding uncovered in the S-work
\cite{dirk} may be stated as follows. Consider a graph code with only one
input and $\left(  n-1\right)  $-output vertices. Its corresponding
coincidence matrix $\Xi_{n\times n}$ can be written as,%
\begin{equation}
\Xi_{n\times n}\overset{\text{def}}{=}\left(
\begin{array}
[c]{cc}%
0_{1\times1} & \mathcal{B}_{1\times\left(  n-1\right)  }^{\dagger}\\
\mathcal{B}_{\left(  n-1\right)  \times\left(  1\right)  } & \mathcal{A}%
_{\left(  n-1\right)  \times\left(  n-1\right)  }%
\end{array}
\right)  \text{,} \label{gammac}%
\end{equation}
where $\mathcal{A}_{\left(  n-1\right)  \times\left(  n-1\right)  }$ denotes
the $\left(  n-1\right)  \times\left(  n-1\right)  $-symmetric adjacency
matrix $\Gamma_{\left(  n-1\right)  \times\left(  n-1\right)  }$. Then, the
graph code with symmetric coincidence matrix $\Xi_{n\times n}$ in Eq.
(\ref{gammac}) is equivalent to stabilizer codes associated with the isotropic
subspace $\mathcal{S}_{\text{isotropic}}$ defined as,%
\begin{equation}
\mathcal{S}_{\text{isotropic}}\overset{\text{def}}{=}\left\{  \left(
\mathcal{A}k\left\vert k\right.  \right)  :k\in\ker\mathcal{B}^{\dagger
}\right\}  \text{,} \label{isoass}%
\end{equation}
that is, omitting phase factors, with the binary stabilizer group
$\mathcal{S}_{\text{binary}}$,%
\begin{equation}
\mathcal{S}_{\text{binary}}\overset{\text{def}}{=}\left\{  g_{k}=X^{k}%
Z^{Ak}:k\in\ker\mathcal{B}^{\dagger}\right\}  \text{.} \label{binaryass}%
\end{equation}
Observe that a stabilizer operator $g_{k}\in\mathcal{S}_{\text{binary}}$ for
an $n$-vertex graph has a $2n$-dimensional binary vector space representation
such that $g_{k}\leftrightarrow v_{g_{k}}\overset{\text{def}}{=}\left(
\mathcal{A}k\left\vert k\right.  \right)  \in\mathbf{F}_{2}^{2n}$.

More generally, consider a $\left[  \left[  n,k,d\right]  \right]  $ binary
quantum stabilizer code associated with a graph $G\left(  V\text{, }E\right)
$ characterized by the $\left(  n+k\right)  \times\left(  n+k\right)  $
symmetric coincidence matrix $\Xi_{\left(  n+k\right)  \times\left(
n+k\right)  }$,%
\begin{equation}
\Xi_{\left(  n+k\right)  \times\left(  n+k\right)  }\overset{\text{def}}%
{=}\left(
\begin{array}
[c]{cc}%
0_{k\times k} & \mathcal{B}_{k\times n}^{\dagger}\\
\mathcal{B}_{n\times k} & \Gamma_{n\times n}%
\end{array}
\right)  \text{.} \label{losai-2}%
\end{equation}
In order to attach the input vertices, $\Xi$ must be constructed such that the
following conditions are satisfied: i) first, $\det\Gamma_{n\times n}=0$
($\operatorname{mod}2$); ii) second,\ the matrix $\mathcal{B}_{k\times
n}^{\dagger}$ must define a $k$-dimensional subspace in $\mathbf{F}_{2}^{n}$
spanned by $k$ linearly independent binary vectors of length $n$ not included
in the \textrm{Span} of the row-vectors\textbf{ }defining the symmetric
adjacency matrix $\Gamma_{n\times n}$,%
\begin{equation}
\text{\textrm{Span}}\left\{  \vec{v}_{1}\text{,..., }\vec{v}_{k}\right\}
\cap\text{ \textrm{Span}}\left\{  \vec{v}_{\Gamma}^{\left(  1\right)
}\text{,..., }\vec{v}_{\Gamma}^{\left(  n\right)  }\text{ }\right\}  =\left\{
\emptyset\right\}  \text{,}%
\end{equation}
where $\vec{v}_{j}\in\mathbf{F}_{2}^{n}$ for $j\in\left\{  1\text{,...,
}k\right\}  $ and $\vec{v}_{\Gamma}^{\left(  i\right)  }\in\mathbf{F}_{2}^{n}$
for $i\in\left\{  1\text{,..., }n\right\}  $; iii) third, Span$\left\{
\vec{v}_{1}\text{,..., }\vec{v}_{k}\right\}  $ contains a vector $\vec{v}%
_{B}\in\mathbf{F}_{2}^{n}$ such that $\vec{v}_{B}\cdot\vec{v}_{\Gamma
}^{\left(  i\right)  }=0$ for any $i\in\left\{  1\text{,..., }n\right\}  $.
Condition i) is needed to avoid disconnected graphs. Condition ii) is required
to have a well defined isometry capable of detecting zero errors. Finally,
condition iii) is needed to generate an isotropic subspace, specifically an
Abelian subgroup of the Pauli group (the so-called stabilizer group) with,%
\begin{equation}
\left(  \Gamma\vec{v}_{\Gamma}^{\left(  l\right)  }\text{, }\vec{v}_{\Gamma
}^{\left(  l\right)  }\right)  \odot\left(  \Gamma\vec{v}_{\Gamma}^{\left(
m\right)  }\text{, }\vec{v}_{\Gamma}^{\left(  m\right)  }\right)  =0\text{, }
\label{28}%
\end{equation}
for any pair $\left(  \vec{v}_{\Gamma}^{\left(  l\right)  }\text{, }\vec
{v}_{\Gamma}^{\left(  m\right)  }\right)  $ with $1\leq l$, $m\leq n$ in
$\left\{  \vec{v}_{\Gamma}^{\left(  1\right)  }\text{,..., }\vec{v}_{\Gamma
}^{\left(  n\right)  }\right\}  $ where the symbol \textquotedblleft$\odot
$\textquotedblright\ denotes the symplectic product \cite{gaitan}. From Eq.
(\ref{28}), we remark that an isotropic subspace of a symplectic vector space
is a vector subspace on which the symplectic form vanishes \cite{robert98}.

Note that in a more general framework like that presented in \cite{dirk},
three types of vertices may be considered: input, auxiliary and output
vertices. The input vertices label the input systems and are used for
encoding. The auxiliary vertices are inputs used as auxiliary degrees of
freedom for implementing additional constraints for the protected code
subspace. Finally, output vertices label the output quantum systems.

\subsection{The Van den Nest-work}

The main achievement of the VdN-work in \cite{bart} is the construction of a
highly effective algorithmic procedure for translating the action of local
Clifford operations on graph states into transformations on their
corresponding graphs. The starting point of the algorithm is a stabilizer
state $\left[  \left[  n,0,d\right]  \right]  $. Moreover, the graphs
considered in the VdN-work have only output vertices and no input vertices.

Before describing this procedure, we remark that it is a\textbf{
}straightforward task to verify that a graph state given by the adjacency
matrix $\Gamma$ corresponds to a stabilizer matrix $\mathcal{S}_{b}%
\overset{\text{def}}{=}\left(  \Gamma\left\vert I\right.  \right)  $ and
transpose stabilizer $\mathcal{T}\overset{\text{def}}{=}\mathcal{S}%
_{b}^{\text{T}}=\binom{\Gamma}{I}$. With\textbf{ }that said, consider a
quantum stabilizer state with stabilizer matrix,%
\begin{equation}
\mathcal{S}_{b}\overset{\text{def}}{=}\left(  Z\left\vert X\right.  \right)
\text{,} \label{ssss}%
\end{equation}
and transpose stabilizer $\mathcal{T}$ given by,%
\begin{equation}
\mathcal{T}\overset{\text{def}}{=}\mathcal{S}_{b}^{\text{T}}=\binom
{Z^{\text{T}}}{X^{\text{T}}}\equiv\binom{A}{B}\text{.} \label{t}%
\end{equation}
Let us focus on the structure of\textbf{ }$\mathcal{S}_{b}$ that
appears\textbf{ }in Eq. (\ref{ssss}). Given a set of generators of the
stabilizer, the stabilizer matrix\textbf{ }$\mathcal{S}_{b}$\textbf{\ }is
generated by assembling the binary representations of the generators as the
rows of a full rank\textbf{ }$\left(  n\times2n\right)  $-matrix. The
transpose of the binary stabilizer matrix (i.e., the transpose
stabilizer)\textbf{ }$\mathcal{T}$\textbf{\ \ }is simply the full rank\textbf{
}$\left(  2n\times n\right)  $-matrix obtained from\textbf{ }$\mathcal{S}_{b}%
$\textbf{\ }upon exchange of rows with columns. The purpose of the algorithmic
scheme is to convert the transpose stabilizer $\mathcal{T}$ in Eq. (\ref{t})
of a given stabilizer state into the transpose stabilizer $\mathcal{T}%
^{\prime}=$ $\binom{A^{\prime}}{B^{\prime}}$ of an equivalent graph state. The
matrix $A^{\prime}$ represents the adjacency matrix of the corresponding
graph. Two scenarios may occur: i) $B$ is an $n\times n$ invertible matrix;
ii) $B$ is not an invertible matrix. In the first scenario where $B$ is
invertible, a right-multiplication of the transpose stabilizer $\mathcal{T}=$
$\binom{A}{B}$ by $B^{-1}$ will implement a basis change, an operation that
generates an equivalent stabilizer state,%
\begin{equation}
\mathcal{T}B^{-1}=\binom{A}{B}B^{-1}=\binom{AB^{-1}}{I}\text{.}%
\end{equation}
Then, the matrix $AB^{-1}$ will denote the resulting adjacency matrix of the
corresponding graph. Furthermore, if the matrix $AB^{-1}$ has nonzero diagonal
elements, we can set these elements to zero in order to satisfy the standard
requirements of the definition of an adjacency matrix of simple graphs. In the
second scenario where $B$ is not invertible, we can always find a suitable
local Clifford unitary transformation $U$ such that \cite{bart},%
\begin{equation}
\mathcal{S}_{b}\overset{\text{def}}{=}\left(  Z\left\vert X\right.  \right)
\overset{U}{\rightarrow}\mathcal{S}_{b}^{\prime}\overset{\text{def}}{=}\left(
Z^{\prime}\left\vert X^{\prime}\right.  \right)  \text{,}%
\end{equation}
and,%
\begin{equation}
\mathcal{T}\overset{\text{def}}{=}\mathcal{S}_{b}^{\text{T}}=\binom
{Z^{\text{T}}}{X^{\text{T}}}\equiv\binom{A}{B}\overset{U}{\rightarrow
}\mathcal{T}^{\prime}\overset{\text{def}}{=}\mathcal{S}_{b}^{\prime\text{T}%
}=\binom{Z^{^{\prime}\text{T}}}{X^{\prime\text{T}}}\equiv\binom{A^{\prime}%
}{B^{\prime}}\text{,}%
\end{equation}
with $\det B^{\prime}\neq0$. Therefore, by right-multiplication of
$\mathcal{T}^{\prime}$ with $B^{\prime-1}$, we obtain%
\begin{equation}
\mathcal{T}^{\prime}B^{\prime-1}=\binom{A^{\prime}}{B^{\prime}}B^{\prime
-1}=\binom{A^{\prime}B^{\prime-1}}{I}\text{.}%
\end{equation}
Thus, the adjacency matrix of the corresponding graph becomes $A^{\prime
}B^{\prime-1}$.

For clarity, we emphasize that the VdN-work was concerned with the special
case of the equivalence between stabilizer and graph states. Instead, it was
Schlingemann in Ref. \cite{dirk} who investigated in a more abstract manner
the equivalence between stabilizer and graph codes. The algorithmic scheme
described above for transforming any binary quantum stabilizer state into a
graph state is very important for our proposed method of generating
graph-theoretic quantum stabilizer codes as will become clear in the following Section.

\section{The procedure}

In this Section, we formally describe our procedure in general terms. Then,
for illustration, we apply it to the graphical construction of the Gottesman
$\left[  \left[  8\text{, }3\text{, }3\right]  \right]  $ eight-qubit quantum
stabilizer code characterized by multi-qubit encoding operators for the error
correction of single amplitude damping errors.

\subsection{Description of the procedure}

Our ultimate goal in this Section is the construction of classical graphs
$G\left(  V\text{, }E\right)  $ with both input and output vertices defined by
the coincidence matrix $\Xi$. The graphs $G\left(  V\text{, }E\right)  $ are
generated for the purpose of verifying the error-correcting capabilities of
the corresponding quantum stabilizer codes via the graph-theoretic error
correction conditions described in the SW-work. In order to achieve this goal,
we propose a systematic procedure based on a rather simple idea: the CWS-,
VdN- and S-works must be combined in such a manner that, in view of our goal,
the weak points of one method should be compensated by the strong points of
another method. A summary of the main graph-theoretic results that we exploit
in our scheme is displayed in Table I.

\begin{table}[t]
\centering
\begin{tabular}
[c]{|c|c|c|c|c|c|}\hline
Year & Authors & Graph-Theoretic Results & Topics for Further Discussion\\\hline
2001 & Schlingemann-Werner & QECCs from graph codes & Non-uniqueness of
representation\\
2002 & Schlingemann & Stabilizer codes as graph codes & Non-uniqueness of
representation\\
2002 & Grassl et \textit{al}. & Stabilizer codes as graph codes &
Non-uniqueness of representation\\
2004 & Van den Nest et \textit{al.} & Local-Clifford equivalence of graph states &
Equivalence of graph codes\\
2009 & Cross et \textit{al}. & Local-Clifford equivalence of graph codes & Construction of
local unitaries\\\hline
\end{tabular}
\caption{Schematic description of years, name of investigators, graph-theoretic results, and issues for further investigation.}%
\end{table}

\subsubsection{Step one}

The CWS formalism represents a general framework wherein both binary-nonbinary
and/or additive-nonadditive quantum codes can be described. For this reason,
the starting point for our procedure is the realization of binary stabilizer
codes as CWS quantum codes. Although this is a relatively straightforward
step, the CWS code that one obtains is not, in general, expressed in the
standard canonical form. From the CWS-work in \cite{cross}, it is known that
there exists local (unitary) Clifford operations that allows, in principle, to
write a CWS\ code that realizes the binary stabilizer code in standard form.
The CWS-work does not however, suggest any algorithmic process by which to
achieve this standard form. In the absence of a systematic process,
identifying a local Clifford unitary $U$ such that $\mathcal{S}_{\text{CWS}%
}^{\prime}\overset{\text{def}}{=}U\mathcal{S}_{\text{CWS}}U^{\dagger}$ (every
element $s^{\prime}\in\mathcal{S}_{\text{CWS}}^{\prime}$ can be written as
$UsU^{\dagger}$ for some $s\in\mathcal{S}_{\text{CWS}}$) may prove to be quite
tedious. Fortunately, this difficulty can be avoided. Before explaining how
this may be\textbf{ }accomplished, let us introduce the codeword stabilizer
matrix $\mathcal{H}_{\mathcal{S}_{\text{CWS}}}\overset{\text{def}}{=}\left(
Z\left\vert X\right.  \right)  $ corresponding to the codeword stabilizer
$\mathcal{S}_{\text{CWS}}$.

\subsubsection{Step two}

The main achievement of the VdN-work in \cite{bart} is the following: each
stabilizer state is shown to be equivalent to a graph state under local
Clifford operations. Observe that a stabilizer state can be regarded as a
quantum code with parameters $\left[  \left[  n\text{, }0\text{, }d\right]
\right]  $. Our idea is to exploit the algorithmic scheme provided by the
VdN-work by translating the starting point of the algorithmic scheme into the
CWS language. To that end, we replace the generator matrix of the stabilizer
state with the codeword stabilizer matrix $\mathcal{H}_{\mathcal{S}%
_{\text{CWS}}}$. The matrix $\mathcal{H}_{\mathcal{S}_{\text{CWS}}}$
corresponds to the codeword stabilizer $\mathcal{S}_{\text{CWS}}$ of the CWS
code that realizes the binary stabilizer code whose graphical depiction is
being sought. In this manner, we may readily apply the VdN algorithmic scheme
so as to determine the standard form of the CWS code and, if necessary, the
explicit expression for the local unitary Clifford operation that connects the
non-standard to the standard forms of the CWS code. After applying the VdN
algorithmic scheme adapted to the CWS formalism, we can generate a graph
specified in terms of a symmetric adjacency matrix $\Gamma$ with only output
vertices. The issue remaining to be addressed is how to attach possible input
vertices to the graph associated with the $\left[  \left[  n\text{, }k\text{,
}d\right]  \right]  $ binary stabilizer codes with $k\neq0$.

\subsubsection{Step three}

Unlike the VdN-work whose findings are limited to binary quantum states, the
S-work extends its range of applicability to include both binary and
non-binary quantum codes. In particular, it was shown in \cite{dirk} that any
stabilizer code is a graph code and vice-versa. Despite the extended range of
applicability of the S-work, it lacked any analogue of the algorithmic scheme
found in the VdN-work. Nevertheless, one result of the S-work provides a
crucial element of our proposed method for generating graph-theoretic quantum
stabilizer codes. Specifically, it was shown that a graph code with associated
graph $G\left(  V\text{, }E\right)  $ containing both input and output
vertices and corresponding symmetric coincidence matrix $\Xi$ is equivalent to
stabilizer codes associated with a suitable isotropic subspace $\mathcal{S}%
_{\text{isotropic}}$. Recall that at the conclusion of step two above, we
essentially have the isotropic subspace and the graph without input vertices
(i.e. the symmetric adjacency matrix $\Gamma$ embedded in the more general
coincidence matrix $\Xi$). Therefore, by exploiting the recently mentioned
finding of the S-work in reverse (we are allowed to do this since a graph code
is equivalent to a stabilizer code and vice-versa), we can construct the full
coincidence matrix $\Xi$ and consequently, be able to attach the input
vertices to the graph. The question now arises:\ what can we do with the
graphical depiction of a binary stabilizer code?

\subsubsection{Step four}

In the SW-work, outstanding graphical QEC conditions were introduced
\cite{werner}. These conditions however, were only partially employed in the
construction of quantum codes associated with graphs and further, the quantum
codes need not have been stabilizer codes. By logically combining the CWS-,
VdN- and S-works, the power of the graphical QEC conditions in \cite{werner}
can be fully exploited in a systematic manner in both directions: from graph
codes to stabilizer codes and vice-versa.

\medskip

In summary, given a binary quantum stabilizer code $\mathcal{C}%
_{\text{stabilizer}}$, the systematic procedure that we propose can be
described in\textbf{ }the following$\ \left(  3+1\right)  $-points. (1)
Realize the stabilizer code $\mathcal{C}_{\text{stabilizer}}$ as a CWS quantum
code $\mathcal{C}_{\text{CWS}}$. (2) Find the graph $G\left(  V\text{,
}E\right)  $ with only output vertices characterized by the symmetric
adjacency matrix $\Gamma$ associated with $\mathcal{C}_{\text{CWS}}$ in the
standard form. This may be accomplished by applying the VdN-work adapted to
the CWS formalism to identify the standard form of the CWS code that realizes
the stabilizer code whose graphical depiction is being sought. (3) Exploit the
S-work to identify the extended graph with both input and output vertices
characterized by the symmetric coincidence matrix $\Xi$ associated with the
isometric encoding map that defines $\mathcal{C}_{\text{CWS}}$. (3+1) Use the
SW-work to apply the graph-theoretic error-correction conditions to the
extended graph in order to explicitly verify the error-correcting capabilities
of the corresponding $\mathcal{C}_{\text{stabilizer}}$ realized as a
$\mathcal{C}_{\text{CWS}}$ quantum code. In Table II, we summarize the
four\textbf{ }steps that characterize our proposed scheme.

\begin{table}[t]
\centering
\begin{tabular}
[c]{|c|c|c|}\hline
Scheme & Technique & Description\\\hline
step-$1$ & CWS-work & Find a codeword stabilizer matrix (CSM) for the $\left[
\left[  n,k,d\right]  \right]  $ stabilizer code\\\hline
step-$2$ & VdN-work & Construct $\Gamma_{n\times n}^{\left(  \mathrm{out}%
\right)  }$ with $\det\Gamma_{n\times n}^{\left(  \mathrm{out}\right)  }=0$
from the standard form of the CSM\\\hline
step-$3$ & S-work & Construct $\Xi_{\left(  n+k\right)  \times\left(
n+k\right)  }^{\left(  \mathrm{in+out}\right)  }\overset{\text{def}}{=}\left(
\begin{array}
[c]{cc}%
O_{k\times k} & \mathcal{B}_{k\times n}^{\dagger\left(  \mathrm{in}\right)
}\\
\mathcal{B}_{n\times k}^{\left(  \mathrm{in}\right)  } & \Gamma_{n\times
n}^{\left(  \mathrm{out}\right)  }%
\end{array}
\right)  $ using $\Gamma_{n\times n}^{\left(  \mathrm{out}\right)  }$ and
identifying $\mathcal{B}_{k\times n}^{\dagger\left(  \mathrm{in}\right)  }%
$\\\hline
step-$4$ & SW-work & Verify the graphical error correction conditions,
$\Xi_{X\cup E}^{I}d^{X\cup E}=0\Rightarrow d^{X\cup E}=0$\\\hline
\end{tabular}
\caption{Schematic description of the four steps that specify our proposed
scheme for constructing a graphical representation of a binary stabilizer
code.}%
\end{table}

\subsection{Application of the procedure}

In our view, there is no better means by which to describe and understand the
effectiveness of our proposed scheme than to work out in detail a simple
illustrative example. In what follows, we wish to determine the graph\textbf{
}associated with the Gottesman $\left[  \left[  8,3,3\right]  \right]  $
stabilizer code, which is a special case of a class of $\left[  \left[
2^{j},2^{j}-j-2,3\right]  \right]  $ codes \cite{danielpra}.

\subsubsection{Step one}

The stabilizer $\mathcal{S}_{\text{b}}^{\text{Gottesman}}$ of the Gottesman
$\left[  \left[  8,3,3\right]  \right]  $ stabilizer code is given by,%
\begin{equation}
\mathcal{S}_{\text{b}}^{\text{Gottesman}}=\left\langle g_{1}\text{, }%
g_{2}\text{, }g_{3}\text{, }g_{4}\text{, }g_{5}\right\rangle \text{,}
\label{sob}%
\end{equation}
where the code encodes three logical qubits into eight physical qubits and
corrects all single-qubit errors. The five stabilizer generators in\ Eq.
(\ref{sob}) are given by \cite{gaitan},%
\begin{align}
&  g_{1}\overset{\text{def}}{=}X^{1}X^{2}X^{3}X^{4}X^{5}X^{6}X^{7}X^{8}\text{,
}g_{2}\overset{\text{def}}{=}Z^{1}Z^{2}Z^{3}Z^{4}Z^{5}Z^{6}Z^{7}Z^{8}\text{,
}g_{3}\overset{\text{def}}{=}X^{2}X^{4}Y^{5}Z^{6}Y^{7}Z^{8}\text{,}\nonumber\\
& \nonumber\\
&  g_{4}\overset{\text{def}}{=}X^{2}Z^{3}Y^{4}X^{6}Z^{7}Y^{8}\text{, }%
g_{5}\overset{\text{def}}{=}Y^{2}X^{3}Z^{4}X^{5}Z^{6}Y^{8}\text{,}%
\end{align}
and a suitable choice for the logical operations $\bar{X}_{i}$ and $\bar
{Z}_{i}$ with $i\in\left\{  1\text{, }2\text{, }3\right\}  $ reads,
\begin{equation}
\bar{X}_{1}\overset{\text{def}}{=}X^{1}X^{2}Z^{6}Z^{8}\text{, }\bar{X}%
_{2}\overset{\text{def}}{=}X^{1}X^{3}Z^{4}Z^{7}\text{, }\bar{X}_{3}%
\overset{\text{def}}{=}X^{1}Z^{4}X^{5}Z^{6}\text{, }\bar{Z}_{1}\overset
{\text{def}}{=}Z^{2}Z^{4}Z^{6}Z^{8}\text{, }\bar{Z}_{2}\overset{\text{def}}%
{=}Z^{3}Z^{4}Z^{7}Z^{8}\text{, }\bar{Z}_{3}\overset{\text{def}}{=}Z^{5}%
Z^{6}Z^{7}Z^{8}\text{.}%
\end{equation}
Therefore, when regarded within the CWS framework \cite{cross}, the
codeword\textbf{ }stabilizer $\mathcal{S}_{\text{CWS}}^{\text{Gottesman}}$ of
the CWS code that realizes the stabilizer code $\mathcal{S}_{\text{b}%
}^{\text{Gottesman}}$ is given by,%
\begin{equation}
\mathcal{S}_{\text{CWS}}^{\text{Gottesman}}\overset{\text{def}}{=}\left\langle
g_{1}\text{, }g_{2}\text{, }g_{3}\text{, }g_{4}\text{, }g_{5}\text{, }\bar
{Z}_{1}\text{, }\bar{Z}_{2}\text{, }\bar{Z}_{3}\right\rangle \text{.}
\label{scws1}%
\end{equation}
From Eq. (\ref{scws1}), we note that the codeword stabilizer matrix
$\mathcal{H}_{\mathcal{S}_{\text{CWS}}^{\text{Gottesman}}}$ associated with
the\textbf{ }codeword stabilizer $\mathcal{S}_{\text{CWS}}^{\text{Gottesman}}$
can be formally written as,
\begin{equation}
\mathcal{H}_{\mathcal{S}_{\text{CWS}}^{\text{Gottesman}}}\overset{\text{def}%
}{=}\left(  Z\left\vert X\right.  \right)  =\left(
\begin{array}
[c]{cccccccc}%
0 & 0 & 0 & 0 & 0 & 0 & 0 & 0\\
1 & 1 & 1 & 1 & 1 & 1 & 1 & 1\\
0 & 0 & 0 & 0 & 1 & 1 & 1 & 1\\
0 & 0 & 1 & 1 & 0 & 0 & 1 & 1\\
0 & 1 & 0 & 1 & 0 & 1 & 0 & 1\\
0 & 1 & 0 & 1 & 0 & 1 & 0 & 1\\
0 & 0 & 1 & 1 & 0 & 0 & 1 & 1\\
0 & 0 & 0 & 0 & 1 & 1 & 1 & 1
\end{array}
\left\vert
\begin{array}
[c]{cccccccc}%
1 & 1 & 1 & 1 & 1 & 1 & 1 & 1\\
0 & 0 & 0 & 0 & 0 & 0 & 0 & 0\\
0 & 1 & 0 & 1 & 1 & 0 & 1 & 0\\
0 & 1 & 0 & 1 & 0 & 1 & 0 & 1\\
0 & 1 & 1 & 0 & 1 & 0 & 0 & 1\\
0 & 0 & 0 & 0 & 0 & 0 & 0 & 0\\
0 & 0 & 0 & 0 & 0 & 0 & 0 & 0\\
0 & 0 & 0 & 0 & 0 & 0 & 0 & 0
\end{array}
\right.  \right)  \text{,}%
\end{equation}
with $\det\left(  X\right)  =0$.

\subsubsection{Step two}

Upon consideration of Eq. (\ref{scws1}), we observe that $\mathcal{S}%
_{\text{CWS}}^{\text{Gottesman}}$ is locally Clifford equivalent to,
\begin{equation}
\mathcal{S}_{\text{CWS}}^{\prime\text{Gottesman}}\overset{\text{def}}%
{=}U\mathcal{S}_{\text{CWS}}^{\text{Gottesman}}U^{\dagger}\text{,}%
\end{equation}
with $U\overset{\text{def}}{=}I^{1}\otimes H^{1}\otimes H^{2}\otimes
H^{3}\otimes H^{5}$ where $H$ denotes the Hadamard transformation. Therefore,
$\mathcal{S}_{\text{CWS}}^{\prime\text{Gottesman}}$ reads\textbf{,}%
\begin{equation}
\mathcal{S}_{\text{CWS}}^{\prime\text{Gottesman}}\overset{\text{def}}%
{=}\left\langle g_{1}^{\prime}\text{, }g_{2}^{\prime}\text{, }g_{3}^{\prime
}\text{, }g_{4}^{\prime}\text{, }g_{5}^{\prime}\text{, }\bar{Z}_{1}^{\prime
}\text{, }\bar{Z}_{2}^{\prime}\text{, }\bar{Z}_{3}^{\prime}\right\rangle
\text{,} \label{15}%
\end{equation}
with,%
\begin{align}
&  g_{1}^{\prime}\overset{\text{def}}{=}Z^{1}Z^{2}Z^{3}X^{4}Z^{5}X^{6}%
X^{7}X^{8}\text{, }g_{2}^{\prime}\overset{\text{def}}{=}X^{1}X^{2}X^{3}%
Z^{4}X^{5}Z^{6}Z^{7}Z^{8}\text{, }g_{3}^{\prime}\overset{\text{def}}{=}%
Z^{2}X^{4}Y^{5}Z^{6}Y^{7}Z^{8}\text{, }\nonumber\\
& \nonumber\\
&  g_{4}^{\prime}\overset{\text{def}}{=}Z^{2}X^{3}Y^{4}X^{6}Z^{7}Y^{8}\text{,
}g_{5}^{\prime}\overset{\text{def}}{=}Y^{2}Z^{3}Z^{4}Z^{5}Z^{6}Y^{8}\text{,}%
\end{align}
and,%
\begin{equation}
\bar{Z}_{1}^{\prime}\overset{\text{def}}{=}X^{2}Z^{4}Z^{6}Z^{8}\text{, }%
\bar{Z}_{2}^{\prime}\overset{\text{def}}{=}X^{3}Z^{4}Z^{7}Z^{8}\text{, }%
\bar{Z}_{3}^{\prime}\overset{\text{def}}{=}X^{5}Z^{6}Z^{7}Z^{8}\text{.}%
\end{equation}
The\textbf{ }codeword stabilizer matrix $\mathcal{H}_{\mathcal{S}_{\text{CWS}%
}^{\prime\text{Gottesman}}}$ associated with the codeword stabilizer
$\mathcal{S}_{\text{CWS}}^{\prime\text{Gottesman}}$ is given by,%
\begin{equation}
\mathcal{H}_{\mathcal{S}_{\text{CWS}}^{\prime\text{Gottesman}}}\overset
{\text{def}}{=}\left(  Z^{\prime}\left\vert X^{\prime}\right.  \right)
=\left(
\begin{array}
[c]{cccccccc}%
1 & 1 & 1 & 0 & 1 & 0 & 0 & 0\\
0 & 0 & 0 & 1 & 0 & 1 & 1 & 1\\
0 & 1 & 0 & 0 & 1 & 1 & 1 & 1\\
0 & 1 & 0 & 1 & 0 & 0 & 1 & 1\\
0 & 1 & 1 & 1 & 1 & 1 & 0 & 1\\
0 & 0 & 0 & 1 & 0 & 1 & 0 & 1\\
0 & 0 & 0 & 1 & 0 & 0 & 1 & 1\\
0 & 0 & 0 & 0 & 0 & 1 & 1 & 1
\end{array}
\left\vert
\begin{array}
[c]{cccccccc}%
0 & 0 & 0 & 1 & 0 & 1 & 1 & 1\\
1 & 1 & 1 & 0 & 1 & 0 & 0 & 0\\
0 & 0 & 0 & 1 & 1 & 0 & 1 & 0\\
0 & 0 & 1 & 1 & 0 & 1 & 0 & 1\\
0 & 1 & 0 & 0 & 0 & 0 & 0 & 1\\
0 & 1 & 0 & 0 & 0 & 0 & 0 & 0\\
0 & 0 & 1 & 0 & 0 & 0 & 0 & 0\\
0 & 0 & 0 & 0 & 1 & 0 & 0 & 0
\end{array}
\right.  \right)  \text{,}%
\end{equation}
where $\det X^{\prime}\neq0$. Therefore, we can find a suitable graph with
only output vertices that is associated with the $\left[  \left[
8,3,3\right]  \right]  $ code by applying the VdN algorithmic procedure. By
considering the transpose of $\mathcal{H}_{\mathcal{S}_{\text{CWS}}^{\prime}}%
$,%
\begin{equation}
\mathcal{T}^{\prime}\overset{\text{def}}{=}\mathcal{H}_{\mathcal{S}%
_{\text{CWS}}^{\prime\text{Gottesman}}}^{\text{T}}=\binom{Z^{\prime\text{T}}%
}{X^{\prime\text{T}}}\equiv\binom{A^{\prime}}{B^{\prime}}\text{,}
\label{bprimo1}%
\end{equation}
we find that $B^{\prime}$ is a $8\times8$ invertible matrix with inverse
$B^{\prime-1}$ given by,%
\begin{equation}
B^{\prime-1}=\left(
\begin{array}
[c]{cccccccc}%
0 & 1 & 0 & 0 & 0 & 0 & 0 & 0\\
0 & 1 & 0 & 0 & 1 & 0 & 0 & 0\\
0 & 1 & 0 & 1 & 0 & 1 & 1 & 0\\
1 & 0 & 1 & 1 & 0 & 0 & 0 & 0\\
0 & 1 & 1 & 0 & 0 & 0 & 0 & 1\\
1 & 0 & 0 & 1 & 0 & 0 & 0 & 0\\
1 & 0 & 1 & 0 & 0 & 0 & 0 & 0\\
1 & 0 & 0 & 1 & 1 & 0 & 0 & 0
\end{array}
\right)  \text{.}%
\end{equation}
Since $\det X^{\prime}\neq0$, we can use the VdN-work to determine the
$8\times8$ adjacency matrix $\Gamma=A^{\prime}B^{\prime-1}$ of a graph that
realizes the $\left[  \left[  8,3,3\right]  \right]  $ code,%
\begin{equation}
\Gamma=A^{\prime}B^{\prime-1}=\left(
\begin{array}
[c]{cccccccc}%
0 & 0 & 0 & 1 & 0 & 1 & 1 & 0\\
0 & 0 & 0 & 1 & 0 & 1 & 0 & 1\\
0 & 0 & 0 & 1 & 0 & 0 & 1 & 1\\
1 & 1 & 1 & 0 & 0 & 0 & 0 & 0\\
0 & 0 & 0 & 0 & 0 & 1 & 1 & 1\\
1 & 1 & 0 & 0 & 1 & 0 & 0 & 0\\
1 & 0 & 1 & 0 & 1 & 0 & 0 & 0\\
0 & 1 & 1 & 0 & 1 & 0 & 0 & 0
\end{array}
\right)  \overset{\text{def}}{=}\Gamma^{\text{Gottesman}}\text{,}
\label{ccube}%
\end{equation}
where $A^{\prime}=Z^{\prime\text{T}}$ and $B^{\prime}=X^{\prime\text{T}}$. We
observe that the graph associated with the adjacency matrix $\Gamma
_{\text{Gottesman}}$ (with $\det\Gamma^{\text{Gottesman}}\neq0$) in Eq.
(\ref{ccube})\ is the cube. As a side remark, we recall that a graph uniquely
determines a graph state, while two graph states determined by two graphs are
equivalent (up to some\emph{\ }local Clifford transformations) if and only if
these two graphs are related to each other via local complementations\emph{\ }%
(LC) \cite{bart}. Avoiding unnecessary formalities, we recall that a local
complementation of a graph on a vertex $v$ can be regarded as the the
operation where, in the neighborhood of $v$, we connect all the disconnected
vertices and disconnect all the connected vertices. In particular, upon
applying a local complementation with respect to the vertex $1$ on the graph
with adjacency matrix $\Gamma_{\text{Gottesman}}$ in Eq. (\ref{ccube}), we
obtain
\begin{equation}
\Gamma^{\text{Gottesman}}\overset{\text{def}}{=}\left(
\begin{array}
[c]{cccccccc}%
0 & 0 & 0 & 1 & 0 & 1 & 1 & 0\\
0 & 0 & 0 & 1 & 0 & 1 & 0 & 1\\
0 & 0 & 0 & 1 & 0 & 0 & 1 & 1\\
1 & 1 & 1 & 0 & 0 & 0 & 0 & 0\\
0 & 0 & 0 & 0 & 0 & 1 & 1 & 1\\
1 & 1 & 0 & 0 & 1 & 0 & 0 & 0\\
1 & 0 & 1 & 0 & 1 & 0 & 0 & 0\\
0 & 1 & 1 & 0 & 1 & 0 & 0 & 0
\end{array}
\right)  \overset{\text{LC}_{v=1}}{\longrightarrow}\Gamma^{\prime
\text{Gottesman}}\overset{\text{def}}{=}\left(
\begin{array}
[c]{cccccccc}%
0 & 0 & 0 & 1 & 0 & 1 & 1 & 0\\
0 & 0 & 1 & 0 & 1 & 0 & 1 & 0\\
0 & 1 & 0 & 0 & 1 & 1 & 0 & 0\\
1 & 0 & 0 & 0 & 1 & 1 & 1 & 1\\
0 & 1 & 1 & 1 & 0 & 0 & 0 & 0\\
1 & 0 & 1 & 1 & 0 & 0 & 1 & 1\\
1 & 1 & 0 & 1 & 0 & 1 & 0 & 1\\
0 & 0 & 0 & 1 & 0 & 1 & 1 & 0
\end{array}
\right)  , \label{833a}%
\end{equation}
where $\det\Gamma^{\prime\text{Gottesman}}=0$.

For the sake of completeness, we also point out that all graphs up to $12$
vertices have been classified under LCs and graph isomorphisms \cite{parker}.
Furthermore, the number of graphs on $n$ unlabeled vertices, or the number of
connected graphs with $n$ vertices, can be found in \cite{sloane}. Finally, a
very recent database of interesting graphs appears in \cite{dam}.

\subsubsection{Step three}

Let us consider the symmetric adjacency matrix $\Gamma_{\text{Gottesman}}$ as
given in Eq. (\ref{ccube}). How do we find the enlarged graph\textbf{ }with
corresponding symmetric coincidence matrix $\Xi_{\text{Gottesman}}$ given
$\Gamma_{\text{Gottesman}}$? Recall that the graph related to $\Gamma
_{\text{Gottesman}}$ realizes a stabilizer code $\mathcal{S}_{\text{b}%
}^{\prime\text{Gottesman}}$ which is locally Clifford equivalent to the
Gottesman $\left[  \left[  8,3,3\right]  \right]  $ code $\mathcal{S}%
_{\text{b}}^{\text{Gottesman}}$ with standard binary stabilizer matrix
$\mathcal{S}_{\text{CWS}}^{\prime\text{Gottesman}}$ given by $\mathcal{S}%
_{\text{CWS}}^{\prime\text{Gottesman}}\overset{\text{def}}{=}\left\langle
g_{1}^{\prime}\text{, }g_{2}^{\prime}\text{, }g_{3}^{\prime}\text{, }%
g_{4}^{\prime}\text{, }g_{5}^{\prime}\text{, }\bar{Z}_{1}^{\prime}\text{,
}\bar{Z}_{2}^{\prime}\text{, }\bar{Z}_{3}^{\prime}\right\rangle $ as defined
in\ Eq. (\ref{15}). Recall further, that for a graph code with both $1$-input
and $n$-output vertices, its corresponding coincidence matrix $\Xi_{\left(
n+1\right)  \times\left(  n+1\right)  }$ has the form expressed in Eq.
(\ref{losai-2}). The graph code with symmetric coincidence matrix
$\Xi_{\left(  n+1\right)  \times\left(  n+1\right)  }$ is equivalent to
stabilizer codes associated\textbf{ }with the isotropic subspace
$\mathcal{S}_{\text{isotropic}}$ defined as $\mathcal{S}_{\text{isotropic}%
}\overset{\text{def}}{=}\left\{  \left(  \mathcal{A}k\left\vert k\right.
\right)  :k\in\ker\mathcal{B}^{\dagger}\right\}  $ in Eq. (\ref{isoass}), that
is, omitting unimportant phase factors, with the binary stabilizer group
$\mathcal{S}_{\text{b}}$ given by $\mathcal{S}_{\text{b}}\overset{\text{def}%
}{=}\left\{  g_{k}=X^{k}Z^{\mathcal{A}k}:k\in\ker\mathcal{B}^{\dagger
}\right\}  $ in Eq. (\ref{binaryass}). In our specific example,\ in agreement
with the three conditions for attaching input vertices as outlined in the
S-work paragraph, we obtain%
\begin{equation}
\Xi^{_{\text{Gottesman}}}=\Xi_{11\times11}^{\text{Gottesman}}\overset
{\text{def}}{=}\left(
\begin{array}
[c]{cc}%
O_{3\times3} & \mathcal{B}_{3\times8}^{\dagger}\\
\mathcal{B}_{8\times3} & \Gamma_{8\times8}^{\prime\text{Gottesman}}%
\end{array}
\right)  \text{,} \label{b0}%
\end{equation}
with $\Gamma_{8\times8}^{\prime\text{Gottesman}}$ in Eq. (\ref{833a}),
$O_{3\times3}$ being the $3\times3$ null-matrix, and $\mathcal{B}_{3\times
8}^{\dagger}$ is given by
\begin{equation}
\mathcal{B}_{3\times8}^{\dagger}\overset{\text{def}}{=}\left(
\begin{array}
[c]{cccccccc}%
1 & 1 & 1 & 0 & 0 & 1 & 1 & 1\\
0 & 1 & 0 & 1 & 1 & 0 & 1 & 0\\
0 & 0 & 1 & 1 & 1 & 1 & 0 & 0
\end{array}
\right)  \text{.} \label{b1}%
\end{equation}
Hence, by applying the S-work and considering $\Gamma^{\prime\text{Gottesman}%
}$ in Eq. (\ref{833a}) together with Eqs. (\ref{b0}) and (\ref{b1}), the
enlarged graph defined by the following $11\times11$ symmetric coincidence
matrix $\Xi^{_{\text{Gottesman}}}$ associated\textbf{\ }with the graph
including both input and output vertices becomes,%
\begin{equation}
\Xi^{_{\text{Gottesman}}}\overset{\text{def}}{=}\left(
\begin{array}
[c]{ccccccccccc}%
0 & 0 & 0 & 1 & 1 & 1 & 0 & 0 & 1 & 1 & 1\\
0 & 0 & 0 & 0 & 1 & 0 & 1 & 1 & 0 & 1 & 0\\
0 & 0 & 0 & 0 & 0 & 1 & 1 & 1 & 1 & 0 & 0\\
1 & 0 & 0 & 0 & 0 & 0 & 1 & 0 & 1 & 1 & 0\\
1 & 1 & 0 & 0 & 0 & 1 & 0 & 1 & 0 & 1 & 0\\
1 & 0 & 1 & 0 & 1 & 0 & 0 & 1 & 1 & 0 & 0\\
0 & 1 & 1 & 1 & 0 & 0 & 0 & 1 & 1 & 1 & 1\\
0 & 1 & 1 & 0 & 1 & 1 & 1 & 0 & 0 & 0 & 0\\
1 & 0 & 1 & 1 & 0 & 1 & 1 & 0 & 0 & 1 & 1\\
1 & 1 & 0 & 1 & 1 & 0 & 1 & 0 & 1 & 0 & 1\\
1 & 0 & 0 & 0 & 0 & 0 & 1 & 0 & 1 & 1 & 0
\end{array}
\right)  \text{.} \label{glb1}%
\end{equation}
For the sake of completeness, we point out that the above-mentioned three
conditions for attaching input vertices are satisfied: i) $\det\Gamma
^{\prime\text{Gottesman}}=0$; ii) $\mathrm{Span}\left\{  \vec{v}_{1}\text{,
}\vec{v}_{2}\text{, }\vec{v}_{3}\right\}  \cap\mathrm{Span}\left\{  \vec
{v}_{\Gamma^{\prime\text{Gottesman}}}^{\left(  1\right)  }\text{,..., }\vec
{v}_{\Gamma^{\prime\text{Gottesman}}}^{\left(  8\right)  }\right\}  =\left\{
\varnothing\right\}  $, with $\mathcal{B}_{8\times3}\overset{\text{def}}%
{=}\left[  \vec{v}_{1}\text{, }\vec{v}_{2}\text{, }\vec{v}_{3}\right]  $ and
$\Gamma^{\prime\text{Gottesman}}\overset{\text{def}}{=}\left[  \vec{v}%
_{\Gamma^{\prime\text{Gottesman}}}^{\left(  1\right)  }\text{,..., }\vec
{v}_{\Gamma^{\prime\text{Gottesman}}}^{\left(  8\right)  }\right]  $; iii)
there exists a $\vec{v}_{\mathcal{B}}\in\mathrm{Span}\left\{  \vec{v}%
_{1}\text{, }\vec{v}_{2}\text{, }\vec{v}_{3}\right\}  $ such that $\vec
{v}_{\mathcal{B}}\cdot\vec{v}_{\Gamma^{\prime\text{Gottesman}}}^{\left(
j\right)  }=0$ for any $1\leq j\leq8$. In Fig. $1$, we display the graph of a
quantum code that is locally Clifford equivalent to Gottesman's $\left[
\left[  8,3,3\right]  \right]  $ code. In Fig. 1, we depict the graph for a
code that is locally Clifford equivalent to the Gottesman $\left[  \left[
8,3,3\right]  \right]  $-code.

\begin{figure}[ptb]
\centering
\includegraphics[width=0.35\textwidth]{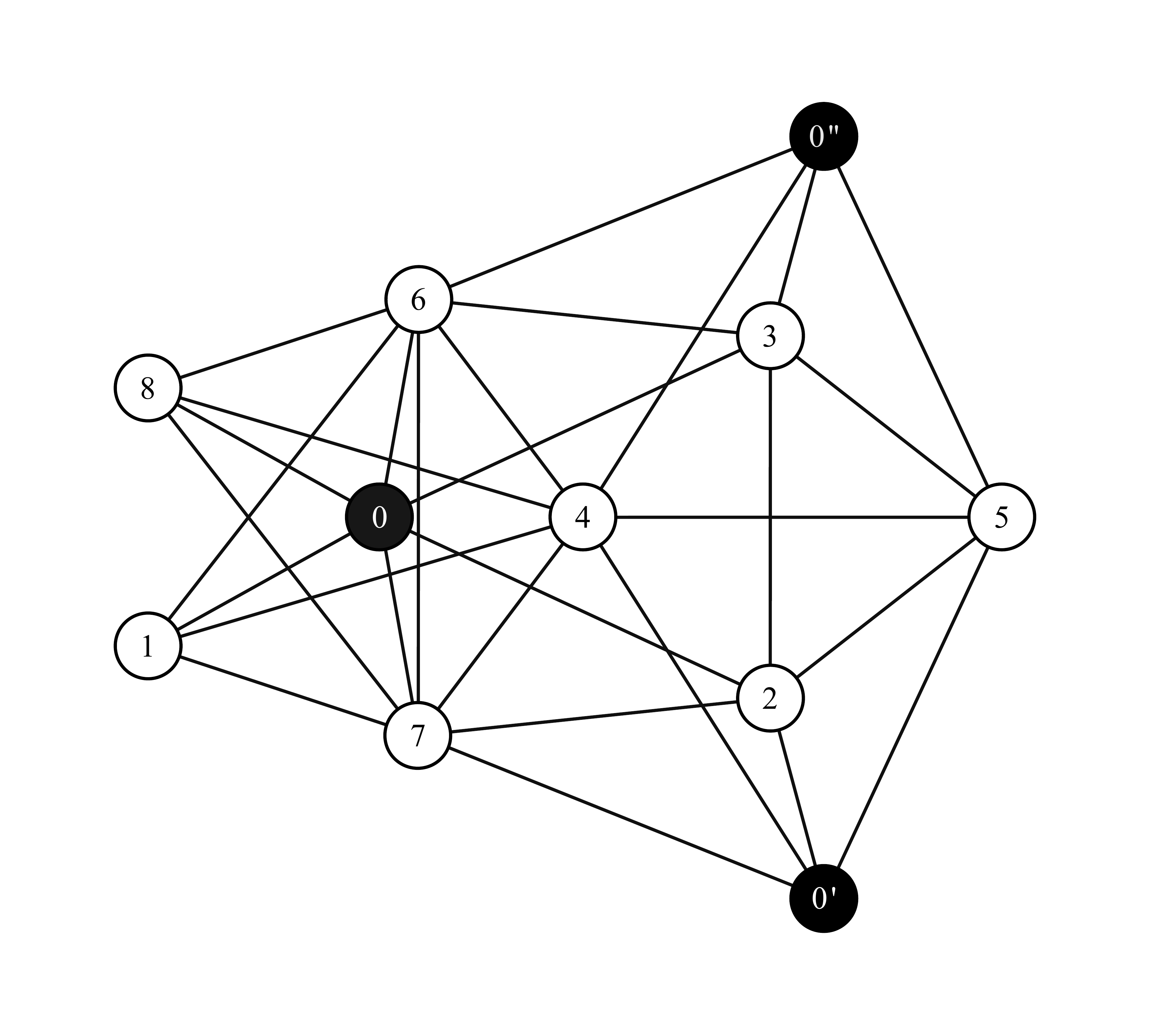}\caption{Graph for a quantum code
that is locally Clifford equivalent to the Gottesman [[8,3,3]]-code.}%
\label{fig1}%
\end{figure}

\subsubsection{Step four}

To verify that the code associated with the $\left[  \left[  8,3,3\right]
\right]  $ graph corrects any one error, we must verify that the code detects
any two errors. Specifically, for every two-element error configuration $E$,
we must have%
\begin{equation}
\Gamma_{X\cup E}^{I}d^{X\cup E}=0\Rightarrow d^{X\cup E}=0\text{,} \label{EC}%
\end{equation}
where $I\overset{\text{def}}{=}Y\backslash E$ with $E$, $X$, $Y$ denoting the
error configuration, the input vertices, and the output vertices,
respectively. In our specific case, we have%
\begin{equation}
X\overset{\text{def}}{=}\left\{  0\text{, }0^{\prime}\text{, }0^{\prime\prime
}\right\}  \text{, }Y\overset{\text{def}}{=}\left\{  1\text{,..., }8\right\}
\text{, }E\overset{\text{def}}{=}\left\{  e_{1}\text{, }e_{2}\right\}  \text{,
and }X\cup E\overset{\text{def}}{=}\left\{  0\text{, }0^{\prime}\text{,
}0^{\prime\prime}\text{, }e_{1}\text{, }e_{2}\right\}  \text{,}%
\end{equation}
where $e_{1,2}\in\left\{  1\text{,..., }8\right\}  $ and the number of
two-element error configurations (that is, the number of sets $X\cup E$) is
$\binom{8}{2}=28$. Observe that when applying Eq. (\ref{EC}), the input
vertices play exactly the same role as an error. For a fixed integration
vertex $y\in I\overset{\text{def}}{=}Y\backslash E$ and a given two-element
error configuration $E$, the condition $\Gamma_{X\cup E}^{I}d^{X\cup E}=0$ is
a set of $\left\vert I\right\vert =\left\vert Y\right\vert -\left\vert
E\right\vert =8-2=6$-equations. In total, we have to verify $28$-systems of
linear algebraic equations with each system defined by $6$-equations. In
particular, for each $y_{\ast}$ we have to sum the $d_{x}$ for all vertices
$x\in X\cup E$. For the sake of clarity, consider the two-error configuration
$E=\left\{  0\text{, }0^{\prime}\text{, }0^{\prime\prime}\text{, }5\text{,
}7\right\}  $. The corresponding linear system specified by six constraint
equations is given by,%
\begin{equation}
\left\{
\begin{array}
[c]{c}%
y_{\ast}=1:d_{0}+d_{7}=0\\
y_{\ast}=2:d_{0}+d_{0^{\prime}}+d_{5}+d_{7}=0\\
y_{\ast}=3:d_{0}+d_{0^{\prime\prime}}+d_{5}=0\\
y_{\ast}=4:d_{0^{\prime}}+d_{0^{\prime\prime}}+d_{5}+d_{7}=0\\
y_{\ast}=6:d_{0}+d_{0^{\prime\prime}}+d_{7}=0\\
y_{\ast}=8:d_{0}+d_{7}=0
\end{array}
\right.  \text{.} \label{system}%
\end{equation}
From simple algebraic manipulations, it can be shown that the system in Eq.
(\ref{system}) yields $d_{0}=d_{0^{\prime}}=d_{0^{\prime\prime}}=d_{5}%
=d_{7}=0$. Therefore, the two-error configurations $\left\{  0\text{,
}0^{\prime}\text{, }0^{\prime\prime}\text{, }5\text{, }7\right\}  $ is
detectable. Following this line of reasoning, it can be verified that any of
the $\binom{8}{2}$ graphical\ two-error configurations $\left\{  0\text{,
}0^{\prime}\text{, }0^{\prime\prime}\text{, }e_{1}\text{, }e_{2}\right\}  $
with $e_{1,2}\in\left\{  1\text{,..., }8\right\}  $ is detectable. Thus, the
code corrects any single-qubit error.

\section{Concluding remarks}

In this article, we proposed a systematic scheme for the construction of
graphs with both input and output vertices associated with arbitrary binary
stabilizer codes. The scheme is characterized by three main steps: first, the
stabilizer code is realized as a CWS quantum code; second, the canonical form
of the CWS\ code is uncovered; third, the input vertices are attached to the
graphs. To verify the effectiveness of the scheme, we implemented the
Gottesman $\left[  \left[  8\text{, }3\text{, }3\right]  \right]  $ stabilizer
code characterized by multi-qubit encoding operators for the error correction
of single amplitude damping errors. In particular, the error-correcting
capabilities of the $\left[  \left[  8\text{, }3\text{, }3\right]  \right]  $
eight-qubit quantum stabilizer code is verified in graph-theoretic terms as
originally advocated by Schlingemann and Werner.

Our two main contributions proposed in this article can be stated as follows:

\begin{itemize}
\item First, our investigation features robust pedagogical and explanatory
value. Indeed, the construction of graphs not only occurs via a novel
alternative scheme, but is also illustrated in a step-by-step fashion via
explicit analytical computations. This step-by-step procedure is helpful for
three reasons. i) First, it highlights one of the main consequence of
the\textbf{ }choices made during explicit application of the scheme, that is
to say, different choices lead to the emergence of non-isomorphic graphs which
yield equivalent graphical quantum codes that are all suitable representations
of the same stabilizer code as originally pointed out by Grassl et \textit{al.
}in Refs. \cite{markus,markus1}. Indeed, while each graph state corresponds
uniquely to a graph, two graph states can be equivalent under local unitary
transformations and yield two, non-isomorphic, distinct graphs. Therefore, as
previously mentioned, there could be non-isomorphic graphs which all yield
graph codes that are distinct suitable representations of the very same
stabilizer code. For an interesting and relatively recent discussion on the
relation between graph states and graph codes, we refer to Ref. \cite{hwang16}%
. ii) Second, it provides a unifying framework to construct graphical
depictions of known binary stabilizer codes, with the former being sparsely
available in the current literature.\textbf{ }During this process, we are also
able to provide new graphical depictions in the presence of both single and
multi-qubit encodings. iii) Third, it allows to reconsider the importance of
the original graphical quantum error correction conditions after
establishing\textbf{ }the clear connection between stabilizer and graph codes.

\item Second, the algorithmic construction of our proposed scheme also offers
pure theoretical and conceptual value. Indeed, we show how the \emph{hybrid
logical blending} of different results- the CWS formalism \cite{cross}, the
algorithmic procedure to transform a stabilizer state into a graph state
\cite{bart}, and the concise link between graph codes and stabilizer codes
\cite{dirk}- lead to an alternative method for constructing graphs of binary
stabilizer codes. In the process, we argue that all these different results
become even more relevant and powerful within our scheme since, among other
things, they solve an old problem in a new fashion. In particular\textbf{,} i)
the Cross et \textit{al}. CWS formalism provides a unifying framework for QEC
code design. Once the stabilizer code is realized as a CWS code, such a code
is locally equivalent to another CWS code in its standard form. In order to
find such a standard form, a suitable local Clifford (unitary) transformation
has to be identified. The CWS formalism does not address this specific issue.
Nevertheless, identifying the standard form of a CWS code can be especially
useful when regarding stabilizer codes as graph codes, and vice-versa. ii) To
determine the standard form of a CWS code corresponding to a given stabilizer
code, we exploit the algorithmic procedure proposed by Van den Nest et
\textit{al}. for transforming any binary quantum stabilizer state into a graph
state. We adapt this algorithmic procedure to the CWS language by replacing
the generator matrix of the stabilizer state with the codeword stabilizer of
the CWS\ code that realizes the binary stabilizer code whose depiction is
being investigated. At the end of this step, one has the adjacency matrix
$\Gamma$ of the graph with only output vertices for the stabilizer code
associated with a suitable isotropic subspace. The work by Van den Nest et
\textit{al}. is not concerned with the issue of finding the extended graph of
the stabilizer code, that is, the graph with both input and output vertices
(to which, we associate the coincidence matrix $\Xi$). iii) To find the
extended graph of a $\left[  \left[  n\text{, }k\text{, }d\right]  \right]  $
binary stabilizer code, we need to attach input vertices to the graph
associated with $\Gamma$. To achieve this goal, we take advantage of a finding
attributed to Schlingemann: a graph code with associated extended graph
including both input and output vertices is equivalent to stabilizer codes
associated with a suitable isotropic subspace. Since each stabilizer code can
be realized as a graph code and vice-versa, we exploit the afore-mentioned
finding in reverse to obtain $\Xi$. iv) After completing the steps described
above, we are able to construct a graph for arbitrary stabilizer codes. Then,
the original graphical quantum error correction and detection conditions
proposed by Schlingemann and Werner in Ref. \cite{werner} can be\textbf{
}re-examined and their usefulness can be tested.
\end{itemize}

The scheme proposed in this article is limited to binary stabilizer codes. How
do we begin addressing from a graph-theoretic perspective non-binary
\cite{beigi2} and continuous-variable (CV) \cite{zang5} codes? What about
non-additive codes? We believe that the extension of our proposed scheme to
codes of this kind will likely pose a highly non-trivial challenge. Despite
its evident limitations, we view our work as a serious effort that not only
expands the interplay among several theoretical formalisms
\cite{werner,dirk,bart,cross}, but it also contributes in a non-trivial manner
to the rich relationship between graph theory and stabilizer formalism in
quantum information \cite{guhne17,adcock20}, with special relevance to quantum
error correction \cite{wagner18}. Finally, in view of the progressive advances
in the field, we are confident that the generalization of our current work
could be readily achieved with concerted and persistent effort.\medskip

\begin{acknowledgments}
C.C. acknowledges helpful discussions on graphs and stabilizer codes with
Yongsoo Hwang, expresses his gratitude to Steven Gassner for technical
assistance with preparing Fig. 1, thanks Sean A.\ Ali for valuable help in
presenting the findings reported in this manuscript in a very clear manner.
Furthermore, C.C. is grateful to Peter van Loock for his guidance during the
preparation of the original and extended (unpublished) version of this work in
Ref. \cite{graph14}. Finally, C. C. acknowledges the ERA-Net CHIST-ERA project
HIPERCOM for financial support for the above mentioned extended (unpublished)
version of this work.
\end{acknowledgments}

\appendix

\section{Technical concepts}

In this Appendix, we present a brief review of several technical concepts.
First, the notions of graphs, graph states and graph codes are introduced.
Second, local Clifford transformations on graph states and local
complementations on graphs are briefly presented.

\subsection{Graphs, graph states, and graph codes}

\emph{Graphs}. A\textbf{\ }graph $G\overset{\text{def}}{=}G\left(  V\text{,
}E\right)  $ is specified by a set $V$ of $n$ vertices together with a set of
edges $E$, the latter being characterized by the adjacency matrix $\Gamma$
\cite{die, west, wilson}. The matrix $\Gamma$ is an $n\times n$ symmetric
matrix with vanishing diagonal elements, where $\Gamma_{ij}=1$ if vertices
$i$, $j$ are connected and $\Gamma_{ij}=0$ otherwise. The neighborhood of a
vertex $i$ corresponds to the set of all vertices $v\in V$ that are connected
to $i$ and is defined by $N_{i}\overset{\text{def}}{=}\left\{  v\in
V:\Gamma_{iv}=1\right\}  $. When the vertices $a$, $b\in V$ coincide with the
end points of an edge, they are said to be adjacent. An $\left\{  a\text{,
}c\right\}  $ path is an ordered list of vertices $a=a_{1}$, $a_{2}$,...,
$a_{n-1}$, $a_{n}=c$, such that $\forall i$, $a_{i}$ and $a_{i+1}$ are
adjacent. A connected graph is one that has an $\left\{  a\text{, }c\right\}
$ path for any two $a$, $c\in V$, otherwise the graph is disconnected. A
vertex represents a physical system, e.g. a qubit (two-dimensional Hilbert
space), qudit ($d$-dimensional Hilbert space, for an example of quantum error
correction with qudits, we refer to \cite{cafaro12} ), or continuous variables
(CV)\ (continuous Hilbert space, for an example of quantum error correction
with CVs, we refer to \cite{ralph11} ). An edge between two vertices
represents the physical interaction between the corresponding systems. In what
follows, we will exclusively consider simple graphs. Simple graphs are those
that contain neither loops (i.e. edges connecting vertices with itself) nor
multiple edges. For the sake of completeness, we remark that is is possible to
make a distinction between different types of vertices. For example, one can
assign some vertices as inputs, and others as outputs.

\emph{Graph states}. Graph states \cite{hans} describe multipartite entangled
states which play a key-role in the graphical construction of QECCs codes and
additionally, play an important role in quantum secret sharing
\cite{damian2008} which to a certain extent, is equivalent to error correction
\cite{anne2013}. For a recent experimental demonstration of a graph state
quantum error correcting code, we refer the reader to \cite{damian2014}.
Consider a system of $n$ qubits that are labeled by the $n$ vertices in $V$.
Furthermore\textbf{, }denote by $I^{i}$, $X^{i}$, $Y^{i}$, $Z^{i}$ (or
equivalently, $X^{i}\equiv\sigma_{x}^{i}$, $Y^{i}\equiv\sigma_{y}^{i}$,
$Z^{i}\equiv\sigma_{z}^{i}$) the identity matrix and the three Pauli operators
acting on the qubit $i\in V$. The $n$-qubit graph state $\left\vert
G\right\rangle $ associated with the graph $G$ is defined by \cite{hein},%
\begin{equation}
\left\vert G\right\rangle \overset{\text{def}}{=}%
%TCIMACRO{\dprod \limits_{\Gamma_{ij}=1}}%
%BeginExpansion
{\displaystyle\prod\limits_{\Gamma_{ij}=1}}
%EndExpansion
\mathcal{U}_{ij}\left\vert +\right\rangle _{x}^{V}=\frac{1}{\sqrt{2^{n}}}%
%TCIMACRO{\dsum \limits_{\vec{\mu}=\mathbf{0}}^{\mathbf{1}}}%
%BeginExpansion
{\displaystyle\sum\limits_{\vec{\mu}=\mathbf{0}}^{\mathbf{1}}}
%EndExpansion
\left(  -1\right)  ^{\frac{1}{2}\vec{\mu}\cdot\Gamma\cdot\vec{\mu}}\left\vert
\vec{\mu}\right\rangle _{z}\text{,}%
\end{equation}
where $\left\vert +\right\rangle _{x}^{V}$ is the joint $+1$ eigenstate of
$X^{i}$ with $i\in V$, and $\mathcal{U}_{ij}$ is the controlled phase gate
between qubits $i$ and $j$ given by,%
\begin{equation}
\mathcal{U}_{ij}\overset{\text{def}}{=}\frac{1}{2}\left[  I+Z_{i}+Z_{j}%
-Z_{i}Z_{j}\right]  \text{,}%
\end{equation}
and $\left\vert \vec{\mu}\right\rangle _{z}$ being the joint eigenstate of
$Z^{i}$ with $i\in V$ and $\left(  -1\right)  ^{\mu_{i}}$ as eigenvalues. The
graph-state basis of the $n$-qubit Hilbert space $\mathcal{H}_{2}^{n}$ is
given by $\left\{  \left\vert G^{C}\right\rangle \overset{\text{def}}{=}%
Z^{C}\left\vert G\right\rangle \right\}  $, where $C$ is an element of the set
of all subsets of $V$ denoted by $2^{V}$. A collection of subsets $\left\{
C_{1}\text{,..., }C_{K}\right\}  $ specifies a $K$-dimensional subspace of
$\mathcal{H}_{2}^{n}$ that is spanned by the graph-state basis $\left\{
\left\vert G^{C_{i}}\right\rangle \right\}  $ with $i=1$,..., $K$. The graph
state $\left\vert G\right\rangle $ is the unique joint $+1$ eigenstate of the
$n$-vertex stabilizers $\mathcal{G}_{i}$ with $i\in V$ defined as \cite{hein},%
\begin{equation}
\mathcal{G}_{i}\overset{\text{def}}{=}X^{i}Z^{N_{i}}\overset{\text{def}}%
{=}X^{i}%
%TCIMACRO{\dprod \limits_{j\in N_{i}}}%
%BeginExpansion
{\displaystyle\prod\limits_{j\in N_{i}}}
%EndExpansion
Z^{j}\text{.}%
\end{equation}
\emph{Graph codes}. A graph code, first introduced in the field of QEC in
\cite{werner} and later reformulated in the graph state formalism \cite{hein},
is defined to be one in which a graph $G$ is given and the codespace (or
coding space) is spanned by a subset of the graph state basis. These states
are regarded as codewords, although we recall that what is significant from
the perspective of the QEC properties is the subspace they span, not the
codewords themselves \cite{robert98}.

\subsection{Local Clifford transformations and local complementations}

\subsubsection{Transformations on quantum states\textbf{\ }}

The Clifford group $\mathcal{C}_{n}$ is the normalizer of the Pauli group
$\mathcal{P}_{\mathcal{H}_{2}^{n}}$ in $\mathcal{U}\left(  2^{n}\right)  $,
i.e. it is the group of unitary operators $U$ satisfying $U\mathcal{P}%
_{\mathcal{H}_{2}^{n}}U^{\dagger}=\mathcal{P}_{\mathcal{H}_{2}^{n}}$. The
local Clifford group $\mathcal{C}_{n}^{l}$ is a subgroup of $\mathcal{C}_{n}$
and is comprised of all $n$-fold tensor products of elements in $\mathcal{C}%
_{1}$. The Clifford group is generated by a simple set of quantum gates,
namely the Hadamard gate $H$, the phase gate $P$ and the CNOT gate
$U_{\text{CNOT}}$ \cite{gaitan}. Using the well-known representations of the
Pauli matrices in the computational basis, it is straightforward to verify
that the action of $H$ on such matrices reads,%
\begin{equation}
\sigma_{x}\rightarrow H\sigma_{x}H^{\dagger}=\sigma_{z}\text{, }\sigma
_{y}\rightarrow H\sigma_{y}H^{\dagger}=-\sigma_{y}\text{, }\sigma
_{z}\rightarrow H\sigma_{z}H^{\dagger}=\sigma_{x}\text{.}%
\end{equation}
The action of the phase gate $P$ on $\sigma_{x}$, $\sigma_{y}$ and $\sigma
_{z}$ is given by,%
\begin{equation}
\sigma_{x}\rightarrow P\sigma_{x}^{\dagger}P=\sigma_{y}\text{, }\sigma
_{y}\rightarrow P\sigma_{y}^{\dagger}P=-\sigma_{x}\text{, }\sigma
_{z}\rightarrow P\sigma_{z}^{\dagger}P=\sigma_{z}\text{.}%
\end{equation}
Finally, the CNOT\ gate leads to the following transformations rules,%
\begin{align}
\sigma_{x}\otimes I  &  \rightarrow U_{\text{CNOT}}\left(  \sigma_{x}\otimes
I\right)  U_{\text{CNOT}}^{\dagger}=\sigma_{x}\otimes\sigma_{x}\text{,
}I\otimes\sigma_{x}\rightarrow U_{\text{CNOT}}\left(  I\otimes\sigma
_{x}\right)  U_{\text{CNOT}}^{\dagger}=I\otimes\sigma_{x}\text{,}\nonumber\\
& \nonumber\\
\sigma_{z}\otimes I  &  \rightarrow U_{\text{CNOT}}\left(  \sigma_{z}\otimes
I\right)  U_{\text{CNOT}}^{\dagger}=\sigma_{z}\otimes I\text{, }I\otimes
\sigma_{z}\rightarrow U_{\text{CNOT}}\left(  I\otimes\sigma_{z}\right)
U_{\text{CNOT}}^{\dagger}=\sigma_{z}\otimes\sigma_{z}\text{.}%
\end{align}
Observe that the CNOT gate propagates bit flip errors from the control to the
target, and phase errors from the target to the control. As a side remark, we
stress that another useful two-qubit gate is the controlled-phase gate
$U_{\text{CP}}\overset{\text{def}}{=}\left(  I\otimes H\right)  U_{\text{CNOT}%
}\left(  I\otimes H\right)  $. The controlled-phase gate has the following
action on the generators of $\mathcal{P}_{\mathcal{H}_{2}^{2}}$,%
\begin{align}
\sigma_{x}\otimes I  &  \rightarrow U_{\text{CP}}\left(  \sigma_{x}\otimes
I\right)  U_{\text{CP}}^{\dagger}=\sigma_{x}\otimes\sigma_{z}\text{, }%
I\otimes\sigma_{x}\rightarrow U_{\text{CP}}\left(  I\otimes\sigma_{x}\right)
U_{\text{CP}}^{\dagger}=\sigma_{z}\otimes\sigma_{x}\text{,}\nonumber\\
& \nonumber\\
\sigma_{z}\otimes I  &  \rightarrow U_{\text{CP}}\left(  \sigma_{z}\otimes
I\right)  U_{\text{CP}}^{\dagger}=\sigma_{z}\otimes I\text{, }I\otimes
\sigma_{z}\rightarrow U_{\text{CP}}\left(  I\otimes\sigma_{z}\right)
U_{\text{CP}}^{\dagger}=I\otimes\sigma_{z}\text{.}%
\end{align}
It is evident that a controlled-phase gate does not propagate phase errors,
though a bit-flip error on one qubit spreads to a phase error to another
qubit. It is worth noting that a unitary operator $U$ that fixes the
stabilizer group $\mathcal{S}_{\text{stabilizer}}$ (we refer to
\cite{daniel-phd} for a detailed exposition of the quantum stabilizer
formalism in QEC) of a quantum stabilizer code $\mathcal{C}_{\text{stabilizer}%
}$ under conjugation is an encoded operation. In other words, $U$ is an
encoded operation that maps codewords to codewords whenever $U\mathcal{S}%
_{\text{stabilizer}}U^{\dagger}=\mathcal{S}_{\text{stabilizer}}$. In
particular, if $\mathcal{S}^{\prime}\overset{\text{def}}{=}U\mathcal{S}%
U^{\dagger}$ (every element of $\mathcal{S}^{\prime}$ can be written as
$UsU^{\dagger}$ for some $s\in\mathcal{S}$) and $\left\vert c\right\rangle $
is a codeword stabilized by every element in $\mathcal{S}$, then $\left\vert
c^{\prime}\right\rangle =U\left\vert c\right\rangle $ is stabilized by every
stabilizer element in $\mathcal{S}^{\prime}$.

\subsubsection{Transformations on graphs}

If there exists a local unitary (LU) transformation $U$ such that $U\left\vert
G\right\rangle =\left\vert G^{\prime}\right\rangle $, the states $\left\vert
G\right\rangle $ and $\left\vert G^{\prime}\right\rangle $ will have the same
entanglement properties. If $\left\vert G\right\rangle $ and $\left\vert
G^{\prime}\right\rangle $ are graph states, then\textbf{ }we say that their
corresponding graphs $G$ and $G^{\prime}$ will represent equivalent quantum
codes, with the same distance, weight distribution, and other properties.
Ascertaining whether two graphs are LU-equivalent is a difficult task, but a
sufficient condition for equivalence was given in \cite{hein}. Let the graphs
$G\overset{\text{def}}{=}G\left(  V\text{, }E\right)  $ and $G^{\prime
}\overset{\text{def}}{=}G^{\prime}\left(  V\text{, }E^{\prime}\right)  $ on
$n$ vertices correspond to the $n$-qubit graph states $\left\vert
G\right\rangle $ and $\left\vert G^{\prime}\right\rangle $. We define the two
$2\times2$ unitary matrices,%
\begin{equation}
\tau_{x}\overset{\text{def}}{=}\sqrt{-i\sigma_{x}}=\frac{1}{\sqrt{2}}\left(
\begin{array}
[c]{cc}%
-1 & i\\
i & -1
\end{array}
\right)  \text{ and }\tau_{z}\overset{\text{def}}{=}\sqrt{i\sigma_{z}}=\left(
\begin{array}
[c]{cc}%
\omega & 0\\
0 & \omega^{3}%
\end{array}
\right)  \text{,}%
\end{equation}
with $\omega^{4}=i^{2}=-1$, where $\sigma_{x}$ and $\sigma_{z}$ are Pauli
matrices. Given a graph $G\overset{\text{def}}{=}G\left(  V=\left\{
0\text{,..., }n-1\right\}  \text{, }E\right)  $ corresponding to the graph
state $\left\vert G\right\rangle $, we define a local unitary transformation
$U_{a}$,%
\begin{equation}
U_{a}\overset{\text{def}}{=}%
%TCIMACRO{\dbigotimes \limits_{i\in N_{a}}}%
%BeginExpansion
{\displaystyle\bigotimes\limits_{i\in N_{a}}}
%EndExpansion
\tau_{x}^{\left(  i\right)  }%
%TCIMACRO{\dbigotimes \limits_{i\notin N_{a}}}%
%BeginExpansion
{\displaystyle\bigotimes\limits_{i\notin N_{a}}}
%EndExpansion
\tau_{z}^{\left(  i\right)  }\text{,}%
\end{equation}
where $a\in V$ is any vertex, $N_{a}\subset V$ is the neighborhood of $a$, and
$\tau_{x}^{\left(  i\right)  }$ means that the transform $\tau_{x}$ should be
applied to the qubit corresponding to vertex $i$. Given a graph $G$, if there
exists a finite sequence of vertices $\left(  u_{0}\text{,..., }%
u_{k-1}\right)  $ such that $U_{u_{k-1}}$...$U_{u_{0}}\left\vert
G\right\rangle =\left\vert G^{\prime}\right\rangle $, then $G$ and $G^{\prime
}$ are LU-equivalent \cite{hein}. It was discovered by Hein et \textit{al}.
and by Van den Nest et \textit{al}. that the sequence of transformations
mapping $\left\vert G\right\rangle $ to $\left\vert G^{\prime}\right\rangle $
can equivalently be expressed as a sequence of simple graph operations taking
$G$ to $G^{\prime}$. In particular, it was shown in \cite{bart} that a graph
$G$ uniquely determines a graph state $\left\vert G\right\rangle $ while two
graph states ($\left\vert G_{1}\right\rangle $ and $\left\vert G_{2}%
\right\rangle $) determined by two graphs ($G_{1}$ and $G_{2}$) are equivalent
up to some local Clifford transformations if and only if these two graphs are
related to each other by local complementations (LCs). The concept of LCs was
originally introduced by Bouchet in \cite{france}. A LC of a graph on a vertex
$v$ refers to the operation in the neighborhood of $v$ whereby\textbf{ }we
connect all disconnected vertices and simultaneously\textbf{ }disconnect all
the connected vertices. All graphs up to $12$ vertices have been classified
under LCs and graph isomorphisms \cite{parker}. It is abundantly clear that
the relationship between graphs and quantum codes can be rather complicated
since one graph may provide non-equivalent codes and different graphs may
provide equivalent codes. It has been established however, that the family of
codes given by a graph is equivalent to the family of codes given by a local
complementation of that graph.

As mentioned earlier, unitary operations $U$ in the local Clifford group
$\mathcal{C}_{n}^{l}$ act on graph states $\left\vert G\right\rangle $. There
also exists graph theoretic rules (i.e. transformations acting on graphs),
which correspond to local Clifford operations. These operations generate the
orbit of any graph state under local Clifford operations. The LC orbit of a
graph $G$ is the set of all non-isomorphic graphs, including $G$ itself, that
can be transformed into $G$ by means of any sequence of LCs and vertex
permutations. The transformation laws for a graph state $\left\vert
G\right\rangle $ and a graph stabilizer $\mathcal{S}_{\Gamma}$ under local
unitary transformations $U$ read,%
\begin{equation}
\left\vert G\right\rangle \rightarrow\left\vert G^{\prime}\right\rangle
=U\left\vert G\right\rangle \text{ and }\mathcal{S}_{\Gamma}\rightarrow
\mathcal{S}_{\Gamma^{\prime}}=U\mathcal{S}_{\Gamma}U^{\dagger}\text{,}
\label{oggi}%
\end{equation}
respectively. Neglecting overall phases, it can be shown that local Clifford
operations $U\in\mathcal{C}_{n}^{l}$ are equivalent to the symplectic
transformations $Q$ of $%
%TCIMACRO{\U{2124} }%
%BeginExpansion
\mathbb{Z}
%EndExpansion
_{2}^{2n}$ which preserve the symplectic inner product \cite{moor}. Therefore,
the $\left(  2n\times2n\right)  $-matrices $Q$ satisfy the relation
$Q^{T}PQ=P$ where the symbol \textquotedblleft$T$\textquotedblright\ denotes
the transpose operation and $P$ is the $\left(  2n\times2n\right)  $-matrix
that defines a symplectic inner product in $%
%TCIMACRO{\U{2124} }%
%BeginExpansion
\mathbb{Z}
%EndExpansion
_{2}^{2n}$, namely%
\begin{equation}
P\overset{\text{def}}{=}\left(
\begin{array}
[c]{cc}%
0 & I\\
I & 0
\end{array}
\right)  \text{.}%
\end{equation}
Furthermore, since local Clifford operations act on each qubit separately,
they have the additional block structure%
\begin{equation}
Q\overset{\text{def}}{=}\left(
\begin{array}
[c]{cc}%
A & B\\
C & D
\end{array}
\right)  \text{,}%
\end{equation}
where the $\left(  n\times n\right)  $-blocks $A$, $B$, $C$, $D$ are diagonal.
It was shown in \cite{bart} that each binary stabilizer state is equivalent to
a graph state. In particular, each graph state characterized by the adjacency
matrix $\Gamma$ corresponds to a stabilizer matrix $\mathcal{S}_{b}%
\overset{\text{def}}{=}\left(  \Gamma\left\vert I\right.  \right)  $ and
transpose stabilizer (generator matrix) $\mathcal{T}\overset{\text{def}}%
{=}\mathcal{S}_{b}^{T}=\binom{\Gamma}{I}$. The generator matrix $\binom
{\Gamma^{\prime}}{I}$ for a graph state with adjacency matrix $\Gamma^{\prime
}$ reads,%
\begin{equation}
\binom{\Gamma}{I}\rightarrow\binom{\Gamma^{\prime}}{I}=\left(
\begin{array}
[c]{cc}%
A & B\\
C & D
\end{array}
\right)  \binom{\Gamma}{I}\left(  C\Gamma+D\right)  ^{-1}\text{,} \label{nons}%
\end{equation}
where,%
\begin{equation}
\Gamma^{\prime}\overset{\text{def}}{=}Q\left(  \Gamma\right)  =\left(
A\Gamma+B\right)  \left(  C\Gamma+D\right)  ^{-1}\text{.} \label{tl1}%
\end{equation}
Observe that in order to have properly defined generator matrices in Eq.
(\ref{nons}), $C\Gamma+D$ must be non-singular and $\Gamma^{\prime}$ must have
vanishing diagonal elements. The graphical analogue of the transformation law
in Eq. (\ref{tl1}) was provided in \cite{bart}. Before stating this result
however, the introduction of additional terminology is required.

Two vertices $i$ and $j$ of a graph $G=G\left(  V\text{, }E\right)  $ are
called adjacent vertices, or neighbors, if $\left\{  i\text{, }j\right\}  \in
E$. The neighborhood $N\left(  i\right)  \subseteq V$ of a vertex $i$ is the
set of all neighbors of $i$. A graph $G^{\prime}=G^{\prime}\left(  V^{\prime
}\text{, }E^{\prime}\right)  $ which satisfies $V^{\prime}\subseteq V$ and
$E^{\prime}\subseteq E$ is a subgraph of $G$ and one writes $G^{\prime
}\subseteq G$. For a subset $A\subseteq V$ of vertices, the induced subgraph
$G\left[  A\right]  \subseteq G$ is the graph with vertex set $A$ and edge set
$\left\{  \left\{  i\text{, }j\right\}  \in E:i\text{, }j\in A\right\}  $. If
$G$ has an adjacency matrix $\Gamma$, its complement $G^{\text{c}}$ is the
graph with adjacency matrix $\Gamma+\mathbf{I}$, where $\mathbf{I}$ is the
$\left(  n\times n\right)  $-matrix whose elements are all ones, except for
the diagonal entries which are zeros. For each vertex $i=1$,..., $n$, a local
complementation $g_{i}$ sends the $n$-vertex graph $G$ to the graph
$g_{i}\left(  G\right)  $ which is obtained by replacing the induced subgraph
$G\left[  N\left(  i\right)  \right]  $ by its complement. In other words,%
\begin{equation}
\Gamma\rightarrow\Gamma^{\prime}\equiv g_{i}(\Gamma)\overset{\text{def}}%
{=}\Gamma+\Gamma\Lambda_{i}\Gamma+\Lambda^{\left(  i\right)  }\text{,}
\label{n1}%
\end{equation}
where $\Lambda_{i}$ has a $1$ on the $i$th diagonal entry, zeros elsewhere and
$\Lambda^{\left(  i\right)  }$ is a diagonal matrix such that it\textbf{
}yields zeros on the diagonal of $g_{i}(\Gamma)$. Finally, the graphical
analogue of Eq. (\ref{tl1}) becomes,%
\begin{equation}
Q_{i}\left(  \Gamma\right)  =g_{i}(\Gamma)\text{,} \label{tl2}%
\end{equation}
with,%
\begin{equation}
Q_{i}\overset{\text{def}}{=}\left(
\begin{array}
[c]{cc}%
I & \text{\textrm{diag}}\left(  \Gamma_{i}\right) \\
\Lambda_{i} & I
\end{array}
\right)  \text{,} \label{n2}%
\end{equation}
where diag$\left(  \Gamma_{i}\right)  \overset{\text{def}}{=}$\textrm{diag}%
$\left(  \Gamma_{i1}\text{,..., }\Gamma_{in}\right)  $. In our proposed
scheme, the matrix $Q_{i}$ in Eq. (\ref{n2}) represents a transformation that
links two distinct binary vector representations $\left\{  v_{S_{k}}\right\}
$ and $\left\{  v_{S_{k}^{\prime}}\right\}  \overset{\text{def}}{=}\left\{
Q_{i}v_{S_{k}}\right\}  $ of operators $\left\{  S_{k}\right\}  $ and
$\left\{  S_{k}^{\prime}\right\}  \overset{\text{def}}{=}\left\{
US_{k}U^{\dagger}\right\}  $ that belong to two distinct codeword stabilizers
$\mathcal{S}_{\text{CWS}}$ and $\mathcal{S}_{\text{CWS}}^{\prime}$ associated
to graphs with adjacency matrices $\Gamma$ and $\Gamma^{\prime}\equiv
g_{i}(\Gamma)$, respectively. Observe that by substituting (\ref{n2}) into
(\ref{tl1}) and using (\ref{n1}), Eq. (\ref{tl2}) gives%
\begin{equation}
Q_{i}\left(  \Gamma\right)  =g_{i}(\Gamma)\Leftrightarrow\Gamma+\Gamma
\Lambda_{i}\Gamma+\Lambda^{\left(  i\right)  }=\Gamma+\Gamma\Lambda_{i}%
\Gamma+\left[  \text{\textrm{diag}}\left(  \Gamma_{i}\right)
+\text{\textrm{diag}}\left(  \Gamma_{i}\right)  \Lambda_{i}\Gamma\right]
\text{,}%
\end{equation}
that is,%
\begin{equation}
\Lambda^{\left(  i\right)  }=\text{\textrm{diag}}\left(  \Gamma_{i}\right)
+\text{\textrm{diag}}\left(  \Gamma_{i}\right)  \Lambda_{i}\Gamma\text{.}%
\end{equation}
The translation of the action of local Clifford operations on graph states
into the action of local complementations on graphs as presented in Eq.
(\ref{tl2}) is a major achievement of \cite{bart}.

\end{document}